\documentclass[pra, final, showpacs, twocolumn, groupedaddress,amsfonts, floatfix]{revtex4}

\usepackage{amsmath, isolatin1, graphicx, color, amsxtra, amssymb, bm}

\newcommand{\ket}[1]{| #1 \rangle}
\newcommand{\bra}[1]{\langle #1 |}
\newcommand{\trace}{\text{Tr}}
\newcommand{\ud}{\text{d}}
\newcommand{\beqn}{\begin{eqnarray}}
\newcommand{\eeqn}{\end{eqnarray}}
\newcommand{\refeqn}[1]{Eq.~(\ref{#1})}
\newcommand{\refeqnn}[1]{Equation~(\ref{#1})}
\newcommand{\reffig}[1]{Fig.~\ref{#1}}
\newcommand{\reffigg}[1]{Figure~\ref{#1}}
\newcommand{\ph}{{\phantom{1}}}
\newcommand{\delphi}{{\Delta\!\!\;\varphi}}

\newcommand{\argmax}{\mathop{\text{arg\,max}}}
\newcommand{\ipspace}{\:\!}
\newcommand{\ts}{\text{s}} 
\newcommand{\ti}{\text{i}} 
\newcommand{\tp}{\text{p}} 
\newcommand{\topt}{\text{opt}}

\begin{document}

\title{Optimal focusing for maximal collection of entangled
  narrow-band\\ photon pairs into single-mode fibers}

\author{Daniel Ljunggren} \email[Corresponding author. Electronic
address: ]{daniellj@kth.se} \homepage{http://www.quantum.se}
\author{Maria Tengner} \affiliation{Department of Microelectronics and
  Information Technology, The Royal Institute of Technology, KTH,
  Electrum 229, SE-164 40 Kista, Sweden}

\date{June 22, 2005}

\begin{abstract}
  We present a theoretical and experimental investigation of the
  emission characteristics and the flux of photon pairs generated by
  spontaneous parametric downconversion in quasi-phase matched bulk
  crystals for the use in quantum communication sources. We show that,
  by careful design, one can attain well defined modes close to the
  fundamental mode of optical fibers and obtain high coupling
  efficiencies also for bulk crystals, these being more easily aligned
  than crystal waveguides.  We distinguish between singles coupling,
  $\gamma_\ts$ and $\gamma_\ti$, conditional coincidence, $\mu_{\ti |
    \ts}$, and pair coupling, $\gamma_\text{c}$, and show how each of
  these parameters can be maximized by varying the focusing of the
  pump mode and the fiber-matched modes using standard optical
  elements.  Specifically we analyze a periodically poled KTP-crystal
  pumped by a 532 nm laser creating photon pairs at 810 nm and 1550
  nm. Numerical calculations lead to coupling efficiencies above
  $93\%$ at optimal focusing, which is found by the geometrical
  relation $L/z_R$ to be $\approx 1$ to 2 for the pump mode and
  $\approx 2$ to 3 for the fiber-modes, where $L$ is the crystal
  length and $z_R$ is the Rayleigh-range of the mode-profile.  These
  results are independent on $L$.  By showing that the single-mode
  bandwidth decreases $\propto 1/L$, we can therefore design the
  source to produce and couple narrow bandwidth photon pairs well into
  the fibers. Smaller bandwidth means both less chromatic dispersion
  for long propagation distances in fibers, and that telecom Bragg
  gratings can be utilized to compensate for broadened photon
  packets---a vital problem for time-multiplexed qubits.  Longer
  crystals also yield an increase in fiber photon flux $\propto
  \sqrt{L}$, and so, assuming correct focusing, we can only see
  advantages using long crystals.
 
\end{abstract}


\maketitle

\section{Introduction}
Spontaneous parametric downconversion (SPDC) accounts for the majority
of entangled photon pairs being produced today. It can be described as
a process in which the electromagnetic field of a single photon
---traveling inside a dielectric material such as a birefringent
crystal---interacts with the atoms by absorption and gives rise to a
nonlinear response in the field of polarization, thereby leaving the
possibility of two or more photons being re-emitted. The laws of
conservation of energy and momentum, together with the randomness and
indistinguishability in the process, also give rise to entanglement, a
nonlocal correlation between the photons.

In quantum communication numerous experiments have been performed to
date involving non-entangled or entangled photons being sent over long
distances, e.g., sources of heralded single photons
\cite{PJF04,FATBBGZ04,AOBT05}, quantum cryptography
\cite{JSWWZ00,NPWBW00,RBGGZ01}, and teleportation \cite{MRTZG03}. A
typical such experiment involves launching each photon of a
(entangled) pair into single-mode fibers and to deliver each one to a
separate party for encoding or decoding.  For successful distribution
over long distances it is vital to have a high rate of pairs generated
at the source, as the attenuation of the fiber is a strongly limiting
factor even at the wavelength of 1550 nm for which the fiber is most
transparent. Today, results with crystals of periodically poled
materials have proved this viable even at moderate pump laser powers
\cite{TTRZBMOG02}, and in some cases the problem has turned into a
matter of limiting the pump power to avoid creating two pairs at the
same time, as this will give false coincidences also when having low
\textit{single}-coupling efficiencies. Instead, what has gained
importance is to have a high \textit{pair}-coupling efficiency that
increases the probability of both photons of a pair being present in
the fibers once they have been created. Furthermore, the use of
time-multiplexed schemes \cite{TBGZ99,FGRZ04} have elicited the need
of launching photons having very narrow frequency bandwidth and long
coherence length in order to limit the effects of dispersion in the
fibers, and to enable the use of interferometers.  Rather than just
filtering the emission at some desired width, as is commonly done, we
will show that it is more efficient in terms of photon-rates to design
the source so that the bandwidth is determined by the crystal length
and fiber coupling alone.

It is the purpose of this article to calculate the maximum coupling
efficiency achievable for photon pairs generated in crystals that are
phase-matched for colinear emission in general, and for periodically
poled $\text{KTiOPO}_4$ (PPKTP) crystals using non-degenerate
quasi-phase matching (QPM) in particular. We look for the optimal
condition for focusing of the pump onto the crystal and focusing of
the emission onto the fiber-end (mode-matching) which maximizes either
the single or the pair-coupling efficiency. The focusing is specified
using the parameter $\xi = L / z_R$, adopted from \cite{BK68} with a
slight modification, where $L$ is the length of the crystal and $z_R$
is the Rayleigh range. We make no thin-crystal approximations, but
take fully into account the focusing geometry of all three interacting
fields: pump, signal, and idler, by decomposing all three fields into
a complete set of orthogonal plane-wave modes.  Other optimizable
parameters of these beams include the direction of the beam axis and
the location of the focus.  Both are regarded fixed, the former being
motivated by the colinear geometry of perfect quasi-phase matching,
and the latter by the fact that focusing onto the center of the
crystal shows to give highest efficiencies.  (Support for the last
claim is given in \cite{KM66} for second harmonic generation.)  We
also regard the center frequency of the beams, the power of the pump,
and the optical properties of the crystal as fixed parameters of the
problem.  We take into account the polychromatic character of the
emission but assume a monochromatic pump (continuous-wave pump), and
we investigate how the coupling efficiency depends on the length of
the crystal and the bandwidth of the wavelength filter in front of the
fiber, but also how the fiber coupling affects the bandwidth of the
coupled photons and the achievable photon-rates.  Our goal is to give
a simple recipe for setting up a colinear source of entangled photon
pairs that optimizes the focusing for the highest single and pair
coupling efficiencies into single mode fibers, and that also
determines a suitable crystal length for a desired bandwidth.

Shortly after the demonstration of parametric generation (PG) and
second harmonic generation (SHG) in the 1960s, Boyd and Kleinman
\cite{BK68}, and others, addressed the focusing in non-colinear
geometries of type-I and showed the importance of optimization for
achieving maximal conversion efficiency in optical parametric
oscillators and frequency doublers.  By using cavities to enhance the
processes one can control the spatial mode of the pump, signal and
idler to support only the fundamental TEM$_{00}$ mode, and under this
condition Boyd and Kleinman suggested that the general optimal
focusing is to set the $\xi$-parameters of all fields the same
(${\xi_\tp = \xi_\ts = \xi_\ti}$).  Later, Guha \textit{et al.}
\cite{GWF82} showed that having unequal parameters can improve the
conversion even further and this is also supported by our results.
The case of type-II SHG have also been studied \cite{Zondy91}, as well
as sum- and difference frequency generation (SFG and DFG)
\cite{Zondy98}, with similar results.  These works were all treating
the light as a classical field, having the signal beam acting as the
relatively strong control-field that is being amplified by the much
stronger pump-beam together with the creation of an idler. It is not
unreasonable to expect that a different situation arises at the
quantum level where both the signal and idler initially are in
uncontrolled vacuum-states.
  
Spontaneous parametric downconversion commonly takes place in bulk
crystal configurations where the signal and idler modes are not
restricted by cavities. This will provide an additional degree of
freedom.  The pump is assumed to be TEM$_{00}$, but the emission will
in general be spatially multimode. A central problem in this article
is to find how much of the emission is in a transverse and
longitudinal fundamental single-mode at different focusing conditions.
For the transverse part, such a single-mode, being Gaussian shaped, is
very close to the Bessel function of the first kind, $J_0(\alpha)$,
which describes the shape of the fundamental fiber mode, and will
therefore provide nearly perfect overlap.  After determining the mode
of the emission we also calculate the $M^2$ factor, commonly used as
a measure of beam-quality, and compare it to experimentally obtained
results.
  
To our knowledge, no analysis has been made to date that characterizes
the colinear emission in quasi-phase matched materials in the way
presented here, i.e., making no assumptions about short crystals or
weak focusing.  It should be noted that the analytical calculations
become difficult without these assumptions and so our goal have been
to formulate the final expression in such a way that it can be
evaluated numerically with relative ease, with only simple assumptions
being made. Taking into account all the needed degrees of
freedom---azimuthal and polar angular spectrum and frequency
included---these numerical computations will become quite
time-consuming on an ordinary personal computer, but still doable.

Various other attempts have been made in the past to characterize the
one- and two-photon spatial optical modes generated by non-colinear
birefringent phase-matching. However, most of them do not use
single-mode fibers to collect photons; Monken \textit{et al.}
\cite{MRP98} and Pittman \textit{et al.} \cite{PSKRSS96} show how
focusing of the pump with a lens can increase the coincidence counts
using an analysis limited to thin crystals, and Aichele \textit{et
  al.}  \cite{ALS02} seek to match the spatio-temporal mode of a
conditionally prepared photon to a classical wave by spectrally and
spatially filtering the trigger, however, without considering focusing
effects.

More recent work connected to ours is a number of papers that consider
the coupling into single-mode fibers; Kurtsiefer \textit{et al.}
\cite{KOW01a} provide, for thin crystals, a hands-on method of
determining the mode of the emission using the relation between the
emission-angle and the wavelength coming from the phase-matching
conditions.  For maximal overlap between the emission-mode and the
fiber-matched mode (target) they presume it is best to choose the
waist of the pump-mode and fiber-matched mode equal. According to
\cite{BK68}, and our results, this is not optimal in general. Bovino
\textit{et al.}  \cite{BVCCGS03} take on a more sophisticated approach
as they carry out the biphoton-state calculation for a non-colinear
source, which takes into account focusing, dispersion, and walk-off
and arrives at a closed expression for the coincidence efficiency.
Other work have been continued along the same lines \cite{CDMW04}; our
conclusion from examining the formulas herein being that high
efficiency can always be achieved for any length of crystal by
choosing the pump waist large enough and the fiber-matched waist small
enough. This is in contrary to our results which show an optimal value
of the focusing parameter ($1 \lesssim \xi \lesssim 3$).  Furthermore,
as shown both in this report and in \cite{BK68}, for a specific
crystal type and wavelength configuration the value of $\xi$ is found
to be a fixed constant for all crystal lengths which makes the
pump-beam waist $w_0^\ph$ relate to the length as ${w_0^\ph \propto
  \sqrt{L}}$ (at optimal focusing), while the results of Ref.
\cite{BVCCGS03, CDMW04} appear to show a linear relationship.  We are
not sure whether these apparent differences are best explained by the
different situations of a non-colinear and colinear source, pulsed vs.
continuous-wave pump, or by otherwise different models or parameters
in either case.  It can be noted that our results seem to provide good
agreement with experiments.

The particular source of photon pairs that spurred the work of this
article is presented by Pelton \textit{et al.} in Ref.
\cite{PMLTKFCL04}. The main idea is to create polarization-entangled
photon pairs at the non-degenerate wavelengths of 810 nm and 1550 nm
from a pump-photon at 532 nm, using two orthogonally oriented
\cite{KWWAE99}, long, bulk KTP crystals. These crystals are
periodically poled for quasi-phase matching which provides colinear
emission suitable for coupling into single-mode fibers, but as told,
also require some optimization for maximum throughput. Preliminary
results can be found in \cite{LTPM04}. Related work is found in
\cite{RBGGZ01, KFMWS04, FMKWS04}

The agenda of this article is as follows. Section \ref{Sec:theMath}
gives a mathematical background, starting in subsection
\ref{Sec:theState} with a review of the one and two-photon state of
the emission derived in Appendix \ref{App:two-photonAmplitude}. In
subsection \ref{Sec:emissionModes} we calculate the emitted modes,
which are qualitatively measured using the beam quality parameter
$M^2$.  This is followed in subsection
\ref{Sec:couplingAndCoincidence} by a mathematical definition of the
single-coupling, coincidence, and pair-coupling efficiencies.  Section
\ref{Sec:theNumerics} presents the numerical results of the coupling
(\ref{Sec:singleCoupling} - \ref{Sec:pairCoupling}), bandwidth
(\ref{Sec:couplingAndBandwidth}), and the $M^2$ factor
(\ref{Sec:M2Coupling}).  Section \ref{Sec:experimental} covers the
experimental setup and the experimental results, where a comparison is
made to numerical predictions.  Conclusions are found in Section
\ref{Sec:conclusions}.

\section{Theoretical description}
\label{Sec:theMath}
\begin{figure}[tb!]
  \begin{center}
    \includegraphics{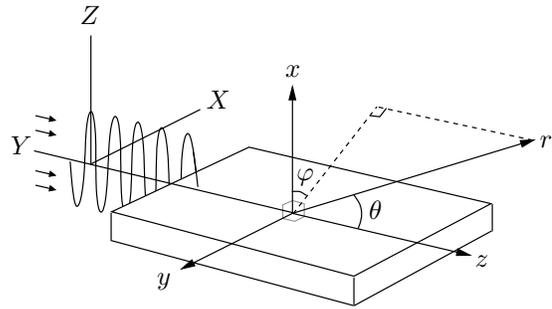}
    \caption{The figure shows the periodically poled crystal with the
      laboratory coordinate system drawn.  Also defined are the
      crystal's axes $X$, $Y$, and $Z$, referring to the polarization
      of the incoming and outgoing electromagnetic fields.}
    \label{Fig:coordinate}
  \end{center}
\end{figure}

The aim of this section is to derive the formulas used for the
numerical calculations of the emission modes, coupling efficiencies,
and emission bandwidths for the emitted quantum state of the SPDC
process, and also to give a physical meaning to these concepts in the
role of single photon sources. We will optimize over the spatial
parameters involved to find the highest quality modes and maximal
coupling efficiencies attainable.  The result is based on a
calculation carried out in Appendix \ref{App:two-photonAmplitude}
involving the Hamiltonian that governs the interaction of spontaneous
parametric downconversion in quasi-phase matched materials. The
crystal is pumped by monochromatic and continuous wave laser light (p)
of frequency $\omega_\tp$, which is propagating in a Gaussian
TEM$_{00}$ mode along the $z$-axis, producing a signal (s) and idler
(i) field in the same direction.  \reffigg{Fig:coordinate} defines the
laboratory axes used; the $z$-axis being along the length $L$ of the
crystal, the $x$-axis along the height, and the $y$-axis along the
width.
The crystal is bi-axial, and the crystal axes $X$, $Y$, and $Z$ are
oriented as shown in the figure. We have chosen the poling period in
the crystal to allow for co-polarized ($Z_\tp Z_\ts Z_\ti$), colinear
down-conversion, but the calculations are general enough to allow
other polarization settings.  The refractive indices, and thus the
phase-matching, depends on the temperature of the crystal and is
determined by the Sellmeier coefficients of \mbox{PPKTP}
\cite{FHHEFBF87, FASR99}. In general we are interested in
phase-matching at non-degenerate wavelengths, and for such cases the
shorter wavelength will be regarded as the signal and the longer
wavelength as the idler.

Many references, following Klyshko \cite{Klyshko88}, start with the
coupled mode equations and look at the evolution of operators to find
the two-photon state from SPDC in terms of a frequency and angular
intensity distribution \cite{MW95}. This is effectively the same as
finding the diagonal elements of the second order moment density
matrix which represent the incoherent part of the information of the
state.  This information is sufficient for determining the shape of
the emission. However, it is not sufficient for determining the
overlap between the emission and a single-mode fiber. In this case we
need the "coherent" information available in the full density matrix.
The approach we take in Appendix \ref{App:two-photonAmplitude} and in
the next subsection is to use the Schrödinger picture and look at
evolution of the state to find the two-photon amplitude. In the
following subsections we then diagonalize the corresponding density
matrix into a sum of coherent parts (eigenmodes), and project each one
onto the fiber-mode so that we can calculate the coupling efficiency
as a sum of overlap coefficients. We also use this decomposition to
calculate the electrical field and beam profile of the emission.

\subsection{\label{Sec:theState}The emitted two-photon state}
The two-photon amplitude describes the joint state of the signal and
idler emission in terms of (internal) angular and frequency spectrum.
Using spherical coordinates (see \reffig{Fig:coordinate}) the
two-photon amplitude derived in \refeqn{Eqn:stateFinal} becomes
\begin{align}
  S(\epsilon, \theta_\ts, \theta_\ti, \delphi) &=
  \frac{4\pi^2 \chi_2^\ph f_1 L}{i\hbar} A^2(\epsilon) \nonumber\\*
  & \times \frac{k_\tp^Z w_{0\tp}^\ph}{\sqrt{2\pi}} e^{-(k_\tp^Z
    w_{0\tp}^\ph)^2 [P^2 + Q^2]/4} \nonumber\\*
  & \times \mathop{\text{sinc}}\left[\frac{L}{2}\Delta k_z^\prime
  \right],
  \label{Eqn:twophotonAmplitude}
\end{align}
where, according to \refeqn{Eqn:deltakPrime}
\begin{align}
  \Delta k_z^\prime &= k_\ts \cos\theta_\ts + k_\ti \cos\theta_\ti -
  k_\tp^Z \sqrt{1 - (P^2 + Q^2)} + K,
  \label{Eqn:A_deltakPrime}
\end{align}
and, according to \refeqn{Eqn:PandQSquared}
\begin{align}
  P^2 + Q^2 &= \nonumber \\
  & \frac{k_\ts^2 \sin^2\theta_\ts + k_\ti^2 \sin^2\theta_\ti + 2 k_\ts k_\ti
    \sin\theta_\ts \sin\theta_\ti \cos(\delphi)}{(k_\tp^Z)^2}.
  \label{Eqn:A_PandQSquared}
\end{align}
All three interacting fields have been decomposed into a complete set
of orthogonal plane-wave modes, $\boldsymbol{k}(\theta, \varphi)$.
The magnitudes of the $k$-vectors, $k_\ts$ and $k_\ti$, are given by
\refeqn{Eqn:kvectorAngleDependence}, $\theta_\ts$ and $\theta_\ti$ are
the internal polar angles of the plane waves of signal and idler
respectively, $\delphi$ is the difference in angle between the
azimuthal angles $\varphi_\ts$ and $\varphi_\ti$, and $\epsilon$ is
the frequency (specified by a single parameter due to exact
energy-matching).  Furthermore, $\chi_2^\ph$ is the nonlinear
coefficient of the crystal, $K$ is the grating constant of the poling,
$L$ is the length of the crystal, and $w_{0\tp}^\ph$ is the pump-beam
waist radius.  $A(\epsilon)$ is the frequency amplitude of the
detector filter having a bandwidth $\Delta\lambda$ ({\small FWHM}) and
a center wavelength $\lambda_c^\ph$ (all wavelengths in vacuum).  Via
the relation $\epsilon = 2\pi c (n_{\!  \scriptscriptstyle \lambda} /
\lambda - n_{\! \scriptscriptstyle \lambda_c^\ph} / \lambda_c^\ph)$
its form, assuming a Gaussian shaped filter, is given by
\begin{align} 
  A(\epsilon; \lambda) = e^{-2\log(2)(\lambda - \lambda_c^\ph)^2 /
    \Delta\lambda^2 }.
  \label{Eqn:frequencyAmplitude}
\end{align}
 
In a plane wave mode-decomposition, \refeqn{Eqn:twophotonAmplitude}
represents the two-photon field (that is generated in the crystal by
the pump field) in the form of a continuous angular spectrum in polar
and azimuthal degrees of freedom. Together with the frequency, the
full state is a tensor-product of four degrees of freedom. We will
need to discretize the spectrum in order to represent it on a
computer. As the size of the Hilbert space of the full ket-vector
becomes very large for a large number of points in resolution, we need
to limit its size to make the numerical calculations feasible. In the
following, the two-photon state is therefore explicitly represented
only by the polar angles of the signal, $\ket{\theta_\ts}$, and the
idler, $\ket{\theta_\ti}$, written as kets, leaving the state
implicitly dependent upon the two remaining degrees of freedom,
$\delphi$ and $\epsilon$.  The purpose of this notation is to reflect
the actual way that the state is numerically implemented as a
one-dimensional array of $\theta$ (the density matrix is a
two-dimensional array), with separate arrays being calculated for each
discrete value $\delphi$ and $\epsilon$. Choosing $N_\theta$ discrete
plane-wave modes as a basis of the polar angle, the two-photon state
can then be formulated as
\begin{align} 
  \ket{\psi_{\ts \ti}^{\delphi, \epsilon}} = \sum_{m,n=1}^{N_\theta}
  S(\epsilon, \theta_{\ts \rule{0pt}{1ex}}^{(m)}, \theta_\ti^{(n)},
  \delphi) \ket{\theta_{\ts \rule{0pt}{1ex}}^{(m)}} \otimes
  \ket{\theta_\ti^{(n)}}.
\end{align}

There are a few approximations that have been made during the
calculation of $S$, apart from the paraxial approximation inherent in
the standard form of the angular spectrum representation of the
Gaussian pump field of \refeqn{Eqn:angularSpectrum}. These include:
\textit{i)} the assumption of a constant pump \mbox{k-vector}
magnitude ${k_\tp = k_\tp^Z}$ in order to remove the implicit
dependence of $\theta_\tp$ and $\varphi_\tp$ in
\refeqn{Eqn:mismatchVectorComponents}, which thus leads to
\refeqn{Eqn:A_PandQSquared}, \textit{ii)} the assumption of an
infinite coherence length of the pump (cw), providing a
$\delta$-function over frequency so that we can describe the signal
and idler by a single frequency $\epsilon$, and \textit{iii)} the
assumption of having the same refractive indices along the crystal's
$X$ and $Y$ axis, such that the $X$-component of the $k$-vectors can
be set to the same as that of $Y$.  The last assumption also provides
a motivation for the output of completely rotationally symmetric
modes, and will greatly simplify the expressions and the numerical
calculations as the azimuthal angle dependence, via $\varphi_\ts$ and
$\varphi_\ti$, is automatically removed from the two-photon amplitude.
The two-photon density matrix is given by
\begin{align}
  \rho_{\ts \ti}^{\delphi, \epsilon} = \ket{\psi_{\ts \ti}^{\delphi,
      \epsilon}}\bra{\psi_{\ts \ti}^{\delphi, \epsilon}},
  \label{Eqn:twoPhotonDensityMatrix}
\end{align}
which now contains four degrees of freedom; $\theta_\ts$ and
$\theta_\ti$ being the two state parameters, and $\delphi$, $\epsilon$
being two other parameters which we will trace over later. Note that
$\rho_{\ts \ti}$ is a description of the emission \textit{inside} the
crystal, not taking into account the refraction between crystal and
air.

\subsection{\label{Sec:emissionModes}The emission modes and the beam
  quality, $M^2$} 

We are interested in the shape of the signal or idler beam profiles
using free detection so that we can compare with images taken by a CCD
camera.  To do this comparison we need to have the beam described in
terms of the electrical field, which is given as the Fourier transform
of the angular spectrum (the density matrix).  The electrical field,
or intensity, then gives the beam profile which, in turn, determines
the $M^2$ factor.

First, each signal or idler beam are made independent of the other
beam by partially tracing over its partner. In the following we trace
over the signal in the polar angle degree of freedom, and in doing so
we get the reduced density matrix for the idler,
\begin{align}
  \rho_\ti^{\delphi,\epsilon} & = \trace_\ts(\rho_{\ts
    \ti}^{\delphi,\epsilon}) = \sum \limits_n^{N_\theta}
  \bra{\theta_s^{(n)}}\rho_{\ts
    \ti}^{\delphi,\epsilon}\ket{\theta_s^{(n)}}.
  \label{Eqn:tracePartner}
\end{align}
The remaining dependence on $\delphi$ can also be removed following
the standard trace-operation, which is here equivalent to a sum over
density matrices,
\begin{align}
  \rho_\ti^{\epsilon} = \trace_{\delphi}(\rho_\ti^{\delphi,\epsilon})
  = \sum \limits_m^{N_\varphi} \rho_\ti^{\delphi_m,\epsilon}.
  \label{Eqn:tracephi}
\end{align}
Additionally, as we could in principle measure the frequency of the
photons at a resolution given by $\Delta\lambda_\text{res} =
\lambda^2/c\Delta t_\text{gate}$ (set by the timing information of the
detectors, $> 1$ ns, to be $< 8$ pm), which generally is much smaller
than the bandwidths of the filters, we need to incoherently sum over
the frequency $\epsilon$ in the same way, giving a final $\rho_i$
describing the state of the idler,
\begin{align}
  \rho_\ti = \trace_{\epsilon}(\rho_\ti^{\epsilon}) = \sum
  \limits_n^{N_\epsilon} \rho_\ti^{\epsilon_n}.
  \label{Eqn:traceepsilon}
\end{align}

\subsubsection*{Mode decomposition}
\begin{figure}[tb!]
  \begin{center}
    \includegraphics[scale = 0.4]{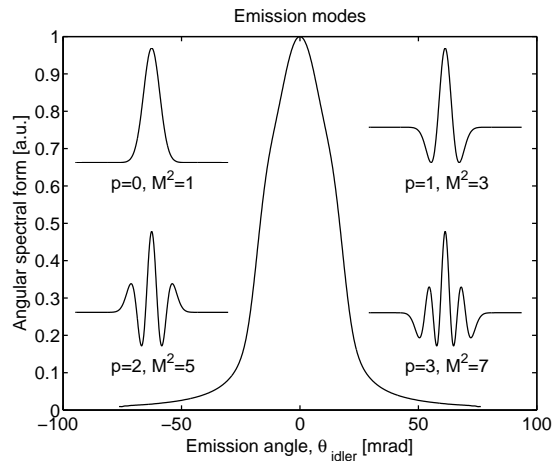}
    \caption{The figure shows an example of the angular spectral form
      $u[\theta_y]$ of the emitted idler light att 1550 nm in a PPKTP
      crystal (central curve) which gives an $M^2$ factor less than 3
      with a filter bandwidth $\Delta\lambda = 10 \text{\ nm}$.  The
      pump at 532 nm is focused close to optimal, $\xi_\tp = 1.3$. The
      insets show the four lowest order LG$_{p0}$ modes which are
      similar, but never the same as the natural eigenmodes of the
      emission, and illustrates how the $M^2$ factor in general grows
      with mode order.}
    \label{Fig:Laguerre}
  \end{center}
\end{figure}

We cannot, however, directly now apply a Fourier transform to the
reduced density matrix $\rho_\ti$, as it is generally mixed.  Instead,
we shall diagonalize $\rho_\ti$ to find its eigenvectors and
eigenvalues.  For such a Hermitian matrix all eigenvalues are real and
the eigenvectors will form a complete orthonormal set. Thence, the set
will represent a natural mode-decomposition of the emission, and
consequently, each vector, or mode, will represent a coherent part of
the emission. The sum of all modes weighted by its corresponding
eigenvalue will determine the state. For each such mode, on the other
hand, we can apply a Fourier transform and thus find the electrical
field modes.  The squared sum of all electrical field modes, again
weighed by the corresponding eigenvalue, will then determine the total
electrical field. We will quantify this to show our future notation;
the reduced density matrix is first diagonalized by $ \bm{T}^{-1} \rho
\bm{T} = \bm{D} \label{Eqn:diagonalization} $, such that $ \bm{T} =
(\ket{\zeta_1}, \ket{\zeta_2}, \ldots, \ket{\zeta_{N_{\theta}}})
\label{Eqn:eigenvectors}$ has the eigenvectors in the
columns, and $\bm{D}$ has the eigenvalues $\lambda_n$ in its diagonal
elements.  The result is a density matrix that can be represented as a
sum of pure states,
\begin{align}
  \rho = \sum \limits_{n = 1}^{N_{\theta}} \lambda_n \ket{\zeta_n}\bra{\zeta_n},
  \label{Eqn:sumofEigenmodes}
\end{align}
where $N_\theta$ is the Hilbert-space dimension.  Following this result, in
\reffig{Fig:Laguerre} is plotted the one-dimensional angular spectral
form $u[\theta_y]$, taken as an integration of the absolute square of
the two-dimensional angular spectral amplitude $a_{xy}[\theta]$. We
have $a_{xy}[\theta] = \sum_n \lambda_n \zeta_n[\theta]$, where
$\zeta_n[\theta]$ is the discrete function representation of
$\ket{\zeta_n}$, and $\theta^2 = \theta_x^2 + \theta_y^2$. Hence,
\begin{align}
  u[\theta_y] = \sum_{\theta_x} \left|a_{xy}\left[\sqrt{\theta_x^2 +
      \theta_y^2}\right]\right|^2,
  \label{Eqn:angularSpectralForm}
\end{align}
is the one-dimensional angular spectral form.

\subsubsection*{The field intensity}
We can now transform the angular spectrum modes $\ket{\zeta_n}$, into
electrical field modes $E_n$. As these modes are rotationally
symmetric and depend on one parameter only, the electrical field is
most suitably expressed through the Hankel transform.  In writing the
transform in the following form we make use of the fact that the
vector $\ket{\zeta_n}$, again written as a discrete function,
$\zeta_n[\theta,\varphi] = \zeta_n[\theta]$, is independent of
$\varphi$. Thus,
\begin{align}
  E_n(x,y,z) &= \sum \limits_{\theta} \lambda_n \zeta_n[\theta]\;
  e^{-i k z \cos\theta} J_0\left(k\sqrt{x^2+y^2}\theta\right),
  \label{Eqn:hankel}
\end{align}
where the basis functions $J_0(\alpha)$ of the Hankel transform are the
Bessel function of zero order and the solution to
$\frac{1}{2\pi}\int_0^{2\pi}\exp{(i \alpha \cos\varphi)} \ud \varphi$.
However, the one-dimensional Fast Hankel Transform (FHT), which would
possibly provide very fast computations, is not widely implemented, at
least not in an efficient form for use in Matlab or Mathematica and
was not available to us at the time for the numerical calculations.
Therefore, the next simplest transform at hand is the two-dimensional
Fourier transform,
\begin{align}
  E_n(x,y,z) &= \sum \limits_{\theta} \sum \limits_{\varphi} \lambda_n
  \zeta_n[\theta,\varphi]\; e^{-i k z \cos\theta} \nonumber \\
  & \times e^{k x \sin\theta \cos\varphi}\; e^{k y \sin\theta
    \sin\varphi}.
  \label{Eqn:fourierLG}
\end{align}
With still two dimensions being used, \refeqn{Eqn:fourierLG} can also
be rewritten using the polar angle components $\theta_x$ and
$\theta_y$,
\begin{align}
  E_n(x,y,z) &= \sum \limits_{\theta_x} \sum \limits_{\theta_y}
  \lambda_n \zeta_n\left[\sqrt{\theta_x^2 + \theta_y^2}\right]\; e^{-i
    k z \cos\left(\sqrt{\theta_x^2 + \theta_y^2}\right)}
  \nonumber \\
  & \times e^{k x \sin\theta_x}\; e^{k y \sin\theta_y},
  \label{Eqn:fourierHG}
\end{align}
where $\theta = \sqrt{\theta_x^2 + \theta_y^2}$.  In this form, which
is the form we will use, \refeqn{Eqn:fourierHG} represents a standard
single two-dimensional FFT.  Note that this transform is, in general,
not separable with respect to $x$ and $y$ into two, but simple, one-dimensional
transforms. This is a characteristic of Laguerre-Gaussian modes and of
the modes emitted by the crystal, in comparison to Hermite-Gaussian
modes which are always separable.

The intensity is now given by incoherently summing all field-modes,
\begin{align}
  I(x,y,z) &= \sum \limits_{n = 1}^{N_\theta} |E_n(x,y,z)|^2.
  \label{Eqn:}
\end{align}
Finally, the transversely integrated intensity profile of the emitted
beam is given by $I(y,z) = \sum_x I(x,y,z)$.

\subsubsection*{Gaussian beam fitting}
The beam waist radius $w(z)$ can be found from the standard deviation
$\sigma(z)$, or the second moment, of the intensity distribution
$I(y,z)$, as $w(z) = 2 \sigma(z)$, see Ref. \cite{Siegman93}. The
standard deviation is known to provide the correct waist estimate for
arbitrary multimode light as opposed to trying to make a curve-fit
with various mode-shapes. Readily, $\sigma^2(z) = \sum_y \left(y -
  \bar{y}(z)\right)^2 I(y,z)$, where ${\bar{y}(z) = \sum_y y I(y,z)}$
is the expectation value with respect to  the spatial position $y$ in the
intensity distribution.  As said, we will use the beam quality factor
$M^2$ to quantify the emission. This factor is determined through the
Rayleigh range
\begin{align}
  z_R^\ph = \frac{\pi w_0^2}{M^2 \lambda},
  \label{Eqn:Rayleigh}
\end{align}
entering the standard Gaussian beam formula
\begin{align}
  w_\text{model}(z) = w_0^\ph \sqrt{1 +
    \left(\frac{z-z_0^\ph}{z_R}\right)^2}.
  \label{Eqn:Gaussian_beam}
\end{align}
By varying the parameters $w_0$ and $M^2$ we can make a curve-fitting
of the model profile $w_\text{model}(z)$ to the actual beam profile
$w(z)$, such that the $M^2$ factor is determined.
\refeqnn{Eqn:Rayleigh} states that the diffraction limited fundamental
Gaussian mode TEM$_{00}$ has a beam quality factor of ${M^2 = 1}$.  As
a comparison, this factor increases for general higher order
Laguerre-Gaussian modes LG$_{pm}$ \cite{Siegman86}, defined by the
radial index $p$ and the azimuthal mode index $m = 0$, such that ${M^2
  = 3}$ for ${p = 1}$, ${M^2 = 5}$ for ${p = 2}$, and ${M^2 = 7}$ for
${p = 3}$ and so on, see \reffig{Fig:Laguerre}.

\subsection{\label{Sec:couplingAndCoincidence}Single coupling,
  coincidence, and pair coupling} 

\begin{figure}[tb!]
  \begin{center}
    \includegraphics{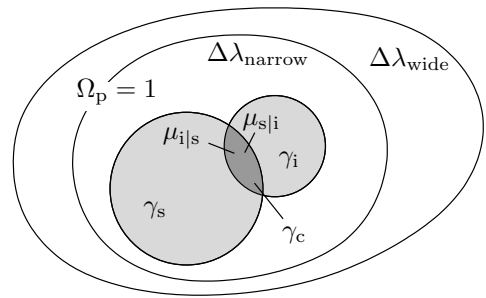}
    \caption{The figure shows a Venn diagram. It illustrates
      the single coupling efficiencies $\gamma_\ts$ and $\gamma_\ti$,
      pair coupling $\gamma_\text{c}$, and conditional coincidences
      $\mu_{\ts |\ti}$ and $\mu_{\ti | \ts}$, which are defined in the
      text. The total amount of pairs $\Omega_\tp$ generated within
      the bandwidth of the detector filter $\Delta\lambda$ is
      normalized to unity, and represents perfect coupling.}
    \label{Fig:vennDiagram}
  \end{center}
\end{figure}

To characterize the source and to optimize the coupling of the
emission into optical fibers we shall make use of three parameters:
single coupling, conditional coincidence, and pair coupling. However,
before we define each of the three coupling parameters we shall
briefly comment on the necessity to relate them to the detection
window being used, i.e., the frequency bandwidth of the detector
filter $\Delta\lambda$. The emission will always fluoresce in a wide
spectrum, and in that sense there is no meaning to speak about a
coupling efficiency for photons that cannot be seen through the window
in any case.  By making a simple normalization to the filter
bandwidth, the coupling probability will consistently measure only how
well photons of specific frequencies are spatially collected into the
fibers. For example, for any fixed filter and no spatial filtering, as
is almost the case with a multimode fiber, and certainly the case in
free-space, the coupling is always perfect.  Effectively, this
normalization enters the calculations through the bandwidth in
\refeqn{Eqn:frequencyAmplitude}.  \reffigg{Fig:vennDiagram} helps to
illustrate the different coupling parameters using a Venn diagram.

\subsubsection*{Single coupling}
The \textit{single-coupling} efficiencies $\gamma_\ts$ and
$\gamma_\ti$ are readily defined as the probability to find a photon
in the fiber which has been emitted within a certain filter bandwidth.
The single-coupling efficiency is useful when maximizing the
individual rate of photons present in the fibers. To calculate the
probability we shall take the overlap of the emitted modes with the
mode of the fiber as seen from the crystal, here called the
\textit{fiber-matched mode}. That is to say, the form of the mode that
can be traced back to the crystal from the fiber-tip, not worrying
about crystal refraction or any other optics in between performing the
actual transformation.  Also, we do not consider any additional
aperture limitations enforced, e.g., by irises.

\begin{figure}[tb!]
  \begin{center}
    \includegraphics{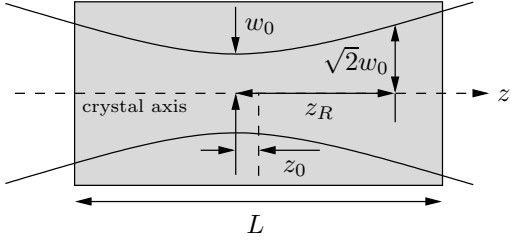}
    \caption{The picture shows the geometry of focusing, with the
      Rayleigh-range $z_R$, the crystal length $L$, the beam waist
      radius $w_0$, and the focus offset $z_0$ being defined.  The
      focusing parameter is defined as $\xi = L/z_R$.}
    \label{Fig:focussing_geometry}
  \end{center}
\end{figure}

The true mode of the fiber is described by a Bessel function.
However, it can be approximated very well with a fundamental Gaussian
which in normalized form is described by
\begin{align}
  \ket{G_{00}} = \frac{k^Z w_{00}^\ph}{\sqrt{2\pi}} e^{-i k^Z
    z_{00}^\ph \cos(\theta) - (k^Z w_{00}^\ph)^2 \sin^2(\theta) / 4}\ 
  \ket{\theta},
  \label{Eqn:fibermode}
\end{align}
where $w_{00}^\ph$ is the beam waist radius of the fiber-matched mode,
TEM$_{00}$, as determined by the focusing system, and $z_{00}^\ph$ is
the location of the corresponding focus (which shall be at the center
of the crystal $z_{00}^\ph = 0$ for optimum coupling), see
\reffig{Fig:focussing_geometry}.

The single-coupling efficiency is trivially given by $\gamma =
\trace(\ket{G_{00}}\bra{G_{00}}\rho)$, but the numerical optimization
converges slowly and badly using this form.  For this reason we shall
exploit the diagonalization and calculate the single coupling
efficiency as the sum of the projection of each emitted mode
$\ket{\zeta_n}$ onto the fiber-matched mode $\ket{G_{00}}$,
\begin{align}
  \gamma = \sum \limits_{n = 1}^{N_\theta} \lambda_n
  |\bra{\zeta_n}\ipspace G_{00}\rangle|^2,
  \label{Eqn:overlap}
\end{align}
where $\ket{\zeta_n}$ is given by the density matrix, $\rho_\ts$ or
$\rho_\ti$, as defined by \refeqn{Eqn:sumofEigenmodes}, resulting in
$\gamma_\ts$ or $\gamma_\ti$ respectively.

\subsubsection*{Optimization}
The maximum achievable coupling efficiency is determined by an
optimization of \refeqn{Eqn:overlap} with respect to the focusing
conditions of either the pump mode, or the fiber-matched signal/idler
mode, or both.  To quantify the focusing we shall use the beam
focusing parameter $\xi = L/z_R$, where $L$ is the length of the
crystal and $z_R$ is the Rayleigh-range (note that we have ${M^2 = 1}$
for both the pump mode and the fiber-matched modes). See
\reffig{Fig:focussing_geometry}. The parameter is suitable as a
dimensionless representation of the focusing geometry.  (As will be
shown further ahead, the results indeed show that the geometry is kept
intact at optimal focusing, irrespectively of the length of the
crystal, which corresponds to a fixed $\xi^\topt$).  In both
\refeqn{Eqn:twophotonAmplitude} and \refeqn{Eqn:fibermode} the
parameter $\xi$ enters through the beam waist radius of the pump mode
$w_{0\tp}$ and the signal/idler fiber-matched mode $w_{00}$, according
to $w_{0\tp} = \sqrt{L \lambda_\tp / \pi \xi_\tp}$, and $w_{00} =
\sqrt{L \lambda_{\ts ,\ti} / \pi \xi_{\ts ,\ti}}$.  We can formalize
the optimization of the signal and idler fiber-matched modes as
\begin{subequations}
  \label{Eqn:optimization_gamma}
  \begin{align}
    \gamma^\topt &= \max_{\xi_{\ts ,\ti}}\ \gamma(\xi_{\tp},
    \xi_{\ts ,\ti}),\\
    \xi^\topt &= \argmax_{\xi_{\ts ,\ti}}\ \gamma(\xi_{\tp},
    \xi_{\ts ,\ti}),
  \end{align}
\end{subequations}
with $\gamma$ given by \refeqn{Eqn:overlap}.

\subsubsection*{Conditional coincidence}
The \textit{conditional coincidences}, $\mu_{\ts |\ti}$ and $\mu_{\ti
  | \ts}$ are useful for the characterization of heralded single
photon sources, and are defined as the probability to find a photon in
either the signal or the idler fiber given that the partner photon has
entered its fiber, whether or not its detected.  The conditional
coincidence probability is found by first projecting the two-photon
amplitude onto the one fiber, and then calculating the overlap with
the other fiber in the same way as for single coupling. In this
example we will search for $\mu_{\ti | \ts}$ and make a conditional
measurement on the signal, defined by the following operator
\begin{align}
  M_\ts = \ket{G_{00}^\text{(s)}}\bra{G_{00}^\text{(s)}}.
  \label{Eqn:condMeasOperator}
\end{align}
Due to the measurement, the derivation of $\rho_\ti$ will be slightly
different here, and we need to take a few steps back and reformulate
the two-photon density matrix $\rho_{\ts \ti}^\epsilon$ as a coherent
sum of amplitudes with respect to $\delphi$, instead of as a
incoherent trace operation in \refeqn{Eqn:tracephi}.  The density
matrix is now written
\begin{align}
  \rho_{\ts \ti}^{\epsilon} = \sum \limits_m \sum \limits_l
  \ket{\psi_{\ts \ti}^{\delphi_m^\ph, \epsilon}}\bra{\psi_{\ts
      \ti}^{\delphi_l^\ph, \epsilon}}.
  \label{Eqn:twoPhotonDensityMatrix2}
\end{align}
Using the measurement operator $M_\ts$, the two-photon density matrix
after the projection becomes
\begin{align}
  \rho_{\ts \ti | \ts}^\epsilon = \frac{M_\ts \otimes \openone_\ti
    \rho_{\ts \ti}^\epsilon M_\ts \otimes \openone_\ti}{\trace(M_\ts
    \otimes \openone_\ti \rho_{\ts \ti}^\epsilon M_\ts \otimes
    \openone_\ti)}.
  \label{Eqn:projection}
\end{align}
The reduced density matrix is readily found by tracing over the
partner, $\rho_{\ti | \ts}^\epsilon = \trace_\ts(\rho_{\ts \ti |
  \ts}^\epsilon)$, which leaves only a trace over frequency,
$\rho_{\ti | \ts} = \sum_n \rho_{\ti | \ts}^{\epsilon_n}$.  The
conditional coincidence is now defined in the same way as for single
coupling; we can replace $\gamma$ by $\mu_{\ti | \ts}$ in
\refeqn{Eqn:overlap}, still using \refeqn{Eqn:sumofEigenmodes} to find
the eigenvalues $\lambda_n$ and eigenmodes $\ket{\zeta_n}$ of
$\rho_{\ti | \ts}$. We have,
\begin{align}
  \mu_{\ti | \ts} = \sum \limits_{n = 1}^{N_\theta} \lambda_n
  |\bra{\zeta_n}\;\!  G^\text{(i)}_{00}\rangle|^2,
  \label{Eqn:conditionalCoupling}
\end{align}
where $\ket{G^\text{(i)}_{00}}$ is the fiber-matched mode of the
idler.  The parameter $\mu_{\ts |\ti}$ follows accordingly, as well as
the formal optimization:
\begin{subequations}
  \label{Eqn:optimization_mu}
  \begin{align}
    \mu^\topt &= \max_{\xi_{\ts ,\ti}}\ 
    \mu(\xi_{\tp}, \xi_{\ts ,\ti}), \\
    \xi^\topt &= \argmax_{\xi_{\ts ,\ti}}\ \mu(\xi_{\tp},
    \xi_{\ts ,\ti}).
  \end{align}
\end{subequations}

\subsubsection*{Pair coupling}
Finally, the \textit{pair-coupling} efficiency $\gamma_\text{c}$ is
defined as the probability to find both photons of a pair in the
respective fiber. This measure tells what fraction of the pairs enters
the fibers compared to the total amount of pairs that are generated
within the frequency bandwidth window. The pair-coupling can be
derived from the single coupling and conditional coincidence using
effectively Bayes's rule, see \reffig{Fig:vennDiagram},
\begin{align}
  \gamma_\text{c} = \mu_{\ti | \ts} \gamma_\ts = \mu_{\ts |\ti}
  \gamma_\ti.
  \label{Eqn:Bayes_gammac}
\end{align}
The alternative is to calculate the coupling via $\gamma_\text{c} =
{\trace(M_\ts \otimes M_\ti\ \rho_{\ts \ti})}$, but this requires the
calculation of $\rho_{\ts \ti}$, which is computationally more
demanding.  When computing $\mu_{\ti | \ts}$ and $\gamma_\ts$ via
\refeqn{Eqn:Bayes_gammac}, using \refeqn{Eqn:conditionalCoupling} and
\refeqn{Eqn:overlap}, the ket is sufficient, because we can simplify
the trace-operation of \refeqn{Eqn:tracePartner}, and also the
projection of \refeqn{Eqn:projection}, to work in ket-space before the
trace over frequency; $\rho_\ti^\epsilon = \trace_\ts(\rho_{\ts
  \ti}^\epsilon) = \sum_{m,n,j} S_{m,j}S^\ast_{n,j}
\ket{\theta_\ti^{(m)}} \bra{\theta_\ti^{(n)}}$. We could also think of
rewriting ${\trace(M_\ts \otimes M_\ti\ \rho_{\ts \ti})}$ using
two-photon kets in the same way, but as $\rho_{\ts \ti}$ generally
becomes a mixture after tracing over frequency this is not an option.
To compute $\gamma_\text{c}$ before the frequency trace is also not an
option numerically, as the trace over frequency involves a for-loop
and optimization performed within it will reduce efficiency heavily.

The measure $\gamma_\text{c}$ should be compared to ${\eta \equiv
  \gamma_\text{c} / \sqrt{\gamma_\ts \gamma_\ti} = \sqrt{\mu_{\ts|\ti}
    \mu_{\ti|\ts}}}$, which is basically $\gamma_\text{c}$ normalized
to $\gamma_\ts$ and $\gamma_\ti$, that have been used by some authors
\cite{BVCCGS03, CDMW04}. The parameter $\eta$ is useful as a type of
measure of correlation that tells how well the focusing system has
been set up to couple the modes of the idler emission to the same as
those conditioned by the signal emission, or vice versa, depending on
which of the two possess the smaller single-coupling efficiency. We
intend to simply plot $\gamma_\text{c}$ as this compares directly to
$\gamma_\ts$ and $\gamma_\ti$ in terms of achievable photon rates; in
principle, $\gamma_\text{c}$ could be low while $\eta$ is high.

\section{Numerical predictions}
\label{Sec:theNumerics}
All results in this section are for the case of a PPKTP crystal with
the poling period $\Lambda = 2\pi/K = 9.6\ \mu\text{m}$ operating at
perfect quasi-phase matching; the pump at 532 nm creates emission at
810 nm and 1550 nm in the absolute forward direction. The temperature
$T = 111~^\circ\text{C}$, which affects the $k$-vector magnitudes, is
chosen such that $k_\tp = k_\ts + k_\ti + K$, see Ref.
\cite{PMLTKFCL04}.

The numerical calculations are implemented in Matlab using
\refeqn{Eqn:twophotonAmplitude}-(\ref{Eqn:A_PandQSquared}). All
refractive indices are determined by the Sellmeier equations
\cite{FHHEFBF87,FASR99}, setting the wavelength and temperature
dependence of the k-vector magnitudes.  The resolution $N_\theta$ of
the discrete angular spectral amplitude representation in the polar
degree are a few hundred points and varies between 1-100 $\mu$radians,
with the higher resolution for short crystals and strong focusing
(wide-spread emission) and the lower resolution for long crystals and
weak focusing (narrow emission). The needed azimuthal angle resolution
$N_\varphi$ is found to be $\gtrsim N_\theta/5$, and the frequency
resolution $N_\epsilon$ varies between a few points for short crystals
to a few hundred points for long crystals where the spectrally induced
contribution to spatial multimode is larger. To spare the computer
from unnecessary workload we observe that the two-photon
density-matrix in \refeqn{Eqn:twoPhotonDensityMatrix} (scaling as
$N_\theta^2$ number of points in size) is always pure and can be fully
represented by its amplitude vector alone (scaling as $N_\theta$), for
all of the calculations.

\subsection{\label{Sec:singleCoupling}Single coupling}
As said earlier, according to our definition the single-coupling
efficiency depends on the emission bandwidth filter that is being
used. This is because of the fact that many of the different
frequencies created in the SPDC process will not couple into a
single-mode fiber.  Looking at a single frequency of the emission, the
angular spectrum of the emission will be described by a single
sinc-function for each of the plane waves of the pump, see
\refeqn{Eqn:twophotonAmplitude}. As will be argued in the next
subsection, most of these sinc-functions will overlap nearly perfectly
at optimal pump-focusing such that the emission is strongly spatially
coherent and define almost a single-mode that will couple well into a
single-mode fiber.  If the pump-focusing is too weak it will create
transverse multimode emission, as the many sinc-functions are then
distributed along the transverse position of the pump beam and do not
coincide. If the pump is instead focused too strongly the effect is
the same, except that the multimode now originates from longitudinal
position, also providing bad coupling.  This is the general picture
using the window of a single emission frequency.

\begin{figure}[tb!]
  \begin{center}
    \includegraphics[scale = 0.4]{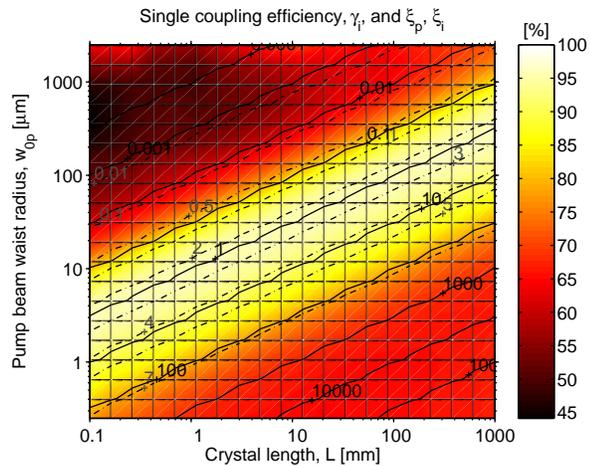}
    \caption{(Color online) The single coupling of the idler
      $\gamma_\ti^\topt$, plotted for a narrow enough filter
      bandwidth, $\Delta\lambda_\text{narrow}$, which shows that about
      95\% of the emission can be coupled into a single-mode fiber at
      optimal focusing.  The solid line shows the pump-focusing
      parameter $\xi_\tp$, and the dashed-dotted lines show the
      focusing of the idler's fiber-matched mode $\xi_\ti^\topt$.  For
      each data sample the idler focusing has been optimized for
      maximum coupling using \refeqn{Eqn:optimization_gamma}.}
    \label{Fig:coupling_vs_crystal_length}
    \end{center}
\end{figure}

\begin{figure}[tb!]
  \begin{center}
    \includegraphics[scale = 0.4]{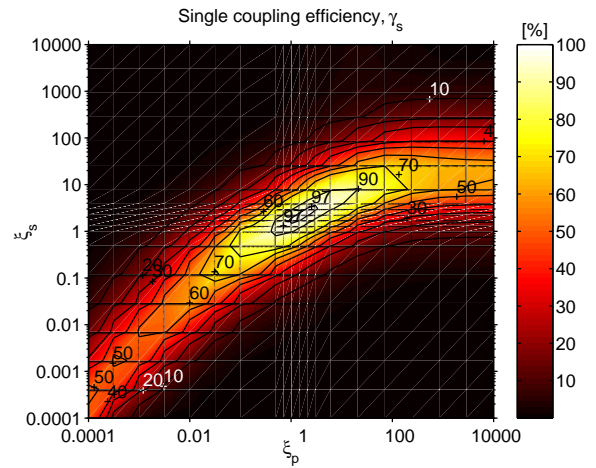}
    \caption{(Color online) The single coupling of the signal $\gamma_\ts$,
      plotted for a narrow enough filter bandwidth,
      $\Delta\lambda_\text{narrow}$, which reaches a maximal 98\% at
      optimal focusing, $\xi_\tp = 1.7$ and $\xi_\ts = 2.3$.}
    \label{Fig:coupling_vs_pumpsignal}
    \end{center}
\end{figure}

\begin{figure}[tb!]
  \begin{center}
    \includegraphics[scale = 0.4]{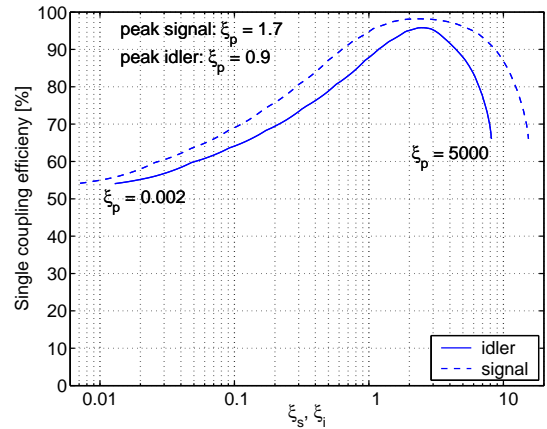}
    \caption{The single couplings, $\gamma_\ti^\topt$ and
      $\gamma_\ts^\topt$, reaches a maximum at $\xi_\tp = 0.9$ for the
      idler, and at $\xi_\tp = 1.7$ for the signal, which corresponds
      to $\xi_\ti^\topt = 2.4$ and $\xi_\ts^\topt = 2.3$. The line
      representing the signal in this graph is essentially a plot of
      the ridge of the surface in
      \reffig{Fig:coupling_vs_pumpsignal}.}
    \label{Fig:coupling_vs_signalidler}
    \end{center}
\end{figure}

If we look at a wide spectrum of the emission, each of the different
frequencies can be seen as composed by a set of sinc-functions, each
set in a different direction, and with every sinc in a set coming from
one plane wave in the decomposition of the pump.  For long crystals,
when the width of the sinc-functions narrows down, the different sets
of sinc-functions will no longer overlap. Within each set the
sinc-functions are spatially well overlapping, thus defining a
coherent single-mode, but as the sets do not overlap the emission will
become spectrally multimode similar to above, also resulting in
spatial multimode. This again provides poor coupling efficiencies.
However, coupling into fibers automatically does some spatial
filtering as it selects only the coherent part of the emission
defining a single-mode, i.e. sinc-functions largely overlapping, and
thereby it also does some frequency filtering. Altogether, this
motivates why we have looked at only a single frequency of the
emission for the results of the numerical calculations of the single
coupling efficiencies shown in
\reffig{Fig:coupling_vs_crystal_length}--\ref{Fig:coupling_vs_signalidler}.
We will refer to this case by saying that we have a "narrow enough"
filter bandwidth, $\Delta\lambda_\text{narrow}$, which maintains a
single-mode at optimal focusing of the pump and the signal and idler
fibers, i.e. the bandwidth is narrow enough that the different
sinc-sets, corresponding to different frequencies, within the
bandwidth overlap (are coherent). Frequency filtering effects, as
those just described, are left to the next section.

\reffigg{Fig:coupling_vs_crystal_length} shows the single-coupling
efficiency of the idler $\gamma_\ti$ plotted against the crystal
length $L$ and the focusing of the pump-beam, via its waist
$w_{0\tp}$. For each sample in the plot, the idler fiber focusing has
been optimized using \refeqn{Eqn:optimization_gamma} to find the
maximum coupling $\gamma_\ti^\topt$. As seen, there is always the same
maximal coupling to be found for any length of the crystal by changing
the pump-beam waist radius accordingly. The straight lines show that
the focusing parameters of both the pump $\xi_\tp$ and the idler fiber
focusing $\xi_\ti^\topt$ are constant, which means that the geometry
of the beam profile and the crystal edges should stay fixed for
different lengths of the crystal for optimal focusing. The said graph
would look nearly the same for the signal emission, and, taking a
different view of the results, \reffig{Fig:coupling_vs_pumpsignal}
clearly shows the importance of choosing the right combination of
focusing for the pump and for the fibers. Interestingly, we observe
that as long as the fiber focusing is matched to the pump focusing,
for any given length of the crystal, then the coupling efficiency will
reach $> 45 \%$ irrespectively of the pump focusing. This fact may
very well explain the relatively high efficiency nevertheless achived
in many fiber-based SPDC-setups for which the experimentalist perhaps
have not worried about changing the pump's focusing, but rather solely
the fiber coupling.

\reffigg{Fig:coupling_vs_signalidler} shows both the signal and idler
coupling in a graph that is parametrized by the pump focusing. In each
case the optimal fiber focusing is found, and plotted along the
horizontal axis.  In this asymmetrical configuration it leads to a
maximal $\gamma_\ts^\topt = 98\%$ when optimizing the focusing for the
810 nm emission ($\xi_\tp = 1.7$ and $\xi_\ts^\topt = 2.3$), and
$\gamma_\ti^\topt = 93\%$ for the 1550 nm emission ($\xi_\tp = 0.9$
and $\xi_\ti^\topt = 2.4$).  The optimal focusing of the pump depends
on the amount of non-degeneracy for each of the wavelengths, e.g., for
the degenerate case (1064 nm) the optimal focusing is $\xi_\tp = 1.4$
and $\xi_{\ts,\ti}^\topt = 2.3$. It should be noted that, in general,
the found optimal focusing parameters do not correspond to a match of
the beam-waist sizes \cite{KOW01a}, but rather to an equal geometry.
However, a matching of the waists are within the same order of
magnitude comparable to using optimal focusing parameters.

\subsection{\label{Sec:pairCoupling}Coincidence and pair coupling}
\begin{figure}[tb!]
  \begin{center}
    \includegraphics[scale = 0.4]{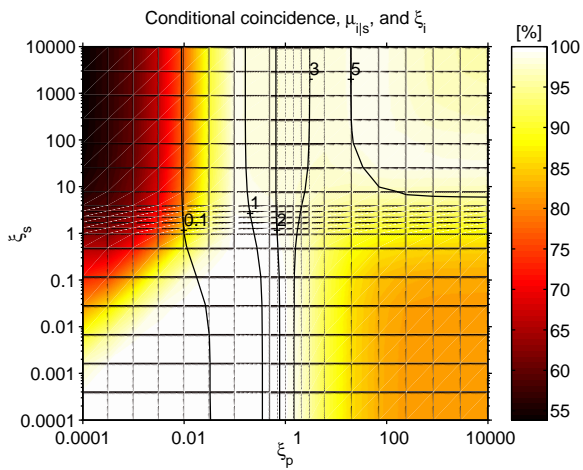}
    \caption{(Color online) The conditional coincidence $\mu_{\ti |
        \ts}$, plotted versus the focusing of the pump $\xi_\tp$ and
      the focusing of the signal's fiber-matched mode $\xi_\ts$. For
      each sample in the graph the focusing of the idler
      ($\xi_\ti^\topt$ = solid lines) is optimized to find the maximum
      $\mu_{\ti | \ts}^\topt$ (up to $100\%$), using
      \refeqn{Eqn:optimization_mu} with a narrow signal filter,
      $\Delta\lambda_\text{narrow}$, and no idler filter.}
    \label{Fig:coincidence_vs_pumpsignal}
  \end{center}
\end{figure}

For any focusing of the pump-beam, the fundamental modes of the signal
and idler emission will be highly correlated, meaning that, e.g., a
signal photon that enters its fiber will have its idler partner
entering the other fiber, provided correct fiber focusing. At optimal
focusing of the pump-beam, this correlation is always high if the
partner beam is focused optimally, independent of the focusing of the
beam that we condition upon. In other words, at optimal focusing of
the pump-beam the conditional coincidence $\mu_{\ti | \ts}$, i.e., the
probability of having the idler photon in the fiber given that the
signal photon is in the fiber, will be mainly set only by its single
coupling probability $\gamma_\ti$, which is always at a high value at
optimal focusing due to the emission being mostly single-mode, see
\reffig{Fig:coincidence_vs_pumpsignal}. 
In contrast, because of the multimode character of the emission at
other pump-beam focusing settings than optimal, a high conditional
coincidence can, in that case, only be attained near optimal focusing
for both the signal and idler fibers.  Each sample in the plot has
been generated using \refeqn{Eqn:optimization_mu} with a narrow
filter, $\Delta\lambda_\text{narrow}$, at the signal side, as defined
earlier, and without a filter at the idler side, when finding the
maximum $\mu_{\ti | \ts}^\topt$ that corresponds to optimal focusing
of the idler, $\xi_\ti^\topt$. As can be deduced from the graph, the
conditional coincidence is always very high, reaching $100 \%$ for
most weaker focusing conditions. When instead using an idler frequency
filter that is matched to the signal filter, then $\mu_{\ti | \ts}$
will be bounded above by $71 \%$, assuming Gaussian shaped filters on
both sides. This limitation follows from the fact that while the
signal photon of a given pair may very well be transmitted through its
filter, the idler may not.  Using \refeqn{Eqn:frequencyAmplitude}, the
maximum number can be easily derived from the normalized overlap
integral $\int |A_s(\epsilon)|^2 |A_i(\epsilon)|^2 \ud \epsilon / \int
|A_s(\epsilon)|^2 \ud \epsilon = 1/\sqrt{2}$, for which we note that
the result is independent of the bandwidth.
\begin{figure}[tb!]
    \begin{center}
      \includegraphics[scale = 0.4]{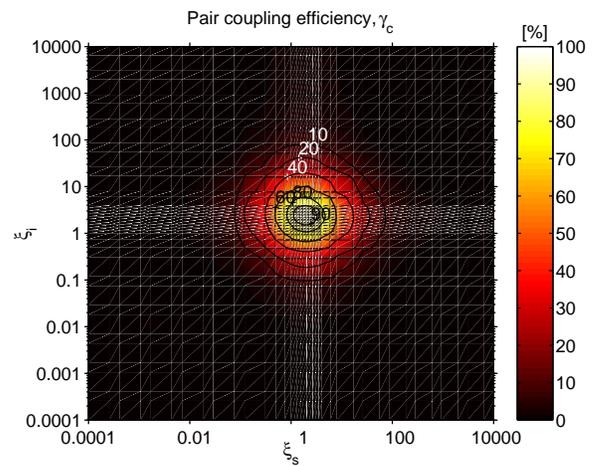}
    \caption{(Color online) The pair coupling $\gamma_\text{c} =
      \mu_{\ti | \ts} \gamma_\ts$ at a pump focusing of $\xi_\tp =
      1.3$, which is trade-off between what is optimal for the signal
      ($\xi_\tp = 1.7$) and the idler ($\xi_\tp = 0.9$) individually.
      At optimal focusing, $\xi_\ts = 2.0$ and $\xi_\ti = 2.3$, the
      maximum $\gamma_\text{c}$ is about $97 \%$, using a narrow
      signal filter, $\Delta\lambda_\text{narrow}$, and no idler
      filter.}
    \label{Fig:pair_coupling_vs_signalidler}
    \end{center}
\end{figure}

Additional qualitative results on the optimal joint focusing can be
found by turning to the pair coupling efficiency $\gamma_\text{c}$. As
opposed to $\mu_{\ti | \ts}$, this measure relates to the total amount
of pairs that is generated, and not only to those conditioned upon. As
shown in \reffig{Fig:pair_coupling_vs_signalidler}, for optimal
pump-beam focusing, there is a maximal value of about $97 \%$ for
$\gamma_\text{c}$ at $\xi_\ts = 2.0$ and $\xi_\ti = 2.3$.  Note that,
since the optimal pump-beam focusing varies for each of the beams for
a non-degenerate wavelength case ($\xi_\tp = 1.7$ for signal and
$\xi_\tp = 0.9$ for idler), we had to find a compromise using $\xi_\tp
= 1.3$. This graph is again plotted using a narrow filter at the
signal and no filter at the idler. \refeqnn{Eqn:Bayes_gammac} tells us
that for matched filters, $\gamma_\text{c}$ will also be limited to
$71 \%$, as long as $\gamma_\ts = 1$ which is achievable with narrow
filters. In general, both the conditional coincidence and the pair
coupling decrease for wide bandwidths; $\mu_{\ti | \ts}$ in such case
being bounded above by $100 \%$ and $\gamma_\text{c}$ bounded above by
the value of $\gamma_\ts$.

In terms of sources of heralded single photons, these results imply
that almost perfect correlation can be achieved by careful focusing
and by having no limiting interference filter on the triggered photon
side; leaving such sources limited entirely by the transmission
imperfections of lenses and filters, and by detector efficiencies.

\subsection{\label{Sec:couplingAndBandwidth}Photon-rate and bandwidth}
\begin{figure}[tb!]
  \begin{center}
    \includegraphics[scale = 0.4]{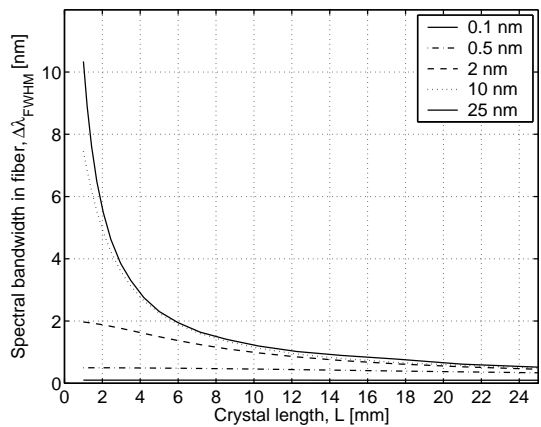}
    \caption{The fiber coupled bandwidth is $\propto 1/L$ for a wide
      enough spectral filter $\Delta\lambda_\text{wide}$, see text,
      which can be said to be the case for the solid line of
      $\Delta\lambda = 25 \text{\ nm}$ for all crystal lengths defined
      by the plot. In the limit of no filter at all, the graph
      corresponds to the single-mode bandwidth
      $\Delta\lambda_\text{SM}$, see \refeqn{Eqn:SMbandwith}. The
      graph shows the result for the signal emission (810 nm) at
      optimal focusing conditions, ${\xi_\tp = 1.7}$ and ${\xi_\ts =
        2.4}$, and the legend shows what filter bandwidth
      $\Delta\lambda$ was used for each line.}
    \label{Fig:bandwidth_vs_crystal_length}
  \end{center}
\end{figure}

\begin{figure}[tb!]
  \begin{center}
    \includegraphics[scale = 0.4]{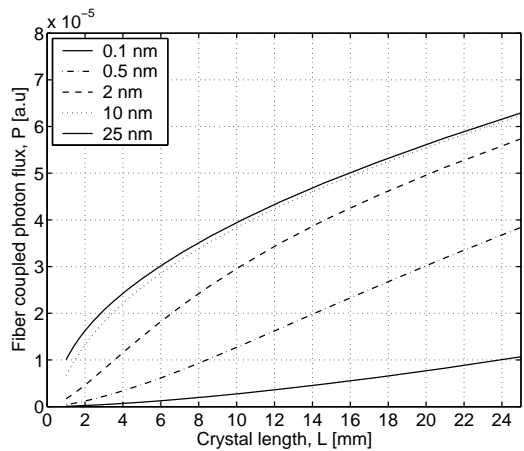}
    \caption{The fiber photon flux is $\propto \sqrt{L}$ for a wide enough
      filter $\Delta\lambda_\text{wide}$, and $\propto L\sqrt{L}$ for
      a narrow enough filter $\Delta\lambda_\text{narrow}$. The filter
      is defined as narrow or wide in relation to the natural
      single-mode bandwidth $\Delta\lambda_\text{SM}$. For the solid
      line of ${\Delta\lambda = 25 \text{\ nm}}$ the case has been
      reached where $\Delta\lambda = \Delta\lambda_\text{wide} >
      \Delta\lambda_\text{SM}$. The graph shows the result for the
      signal emission (810 nm) at optimal focusing conditions,
      ${\xi_\tp = 1.7}$ and ${\xi_\ts = 2.4}$, and the legend shows
      what filter bandwidth $\Delta\lambda$ was used for each line.}
    \label{Fig:flux_vs_crystal_length}
  \end{center}
\end{figure}

In this subsection we will look at the achievable photon fluxes in
free-space and in single-mode fibers and its dependence on the crystal
length. As we will argue, and we have shown numerically, this
dependence will in turn depend on the chosen frequency filter.  Our
arguments will follow a series of steps, where the later steps include
the effects of spatial and spectral filtering. The final results are
found in \reffig{Fig:bandwidth_vs_crystal_length} and
\reffig{Fig:flux_vs_crystal_length}.

As a first step, imagine the pump beam to be a single plane wave that
is perfectly phase-matched for a single frequency of the signal and
the idler along the $z$-axis, called here the forward direction. In
this case, by looking at the two-photon amplitude
\refeqn{Eqn:twophotonAmplitude}, we see that the height of the
sinc-function, which describes the angular spectrum, is $\propto L$,
corresponding to an $L^2$ dependence for the intensity. (One should
imagine two-dimensional, ``mexican-hat-like'', sinc-functions.) The
width of the sinc will shrink $\propto 1/L$, such that the flux will
increase $\propto L$. This argument is still valid considering the
spatial transverse multimode emission created by such a plane wave
pump, discussed earlier.

As a second step, consider a focused pump being composed of many
differently directed plane waves. In this case, still looking at the
same single frequency emitted, each such plane wave will phase-match a
little less strongly than the one in the absolute forward direction.
We will have a collection of sinc-functions being added together, each
originating from a different plane pump wave, and numerical
calculations show that the combined total width, or envelope, of these
sinc-functions will decrease for longer crystals, thus adding to the
previous result a factor $1/\sqrt{L}$, with the flux now becoming
$\propto \sqrt{L}$.

The third step includes the observation that the energy of the pump
beam is concentrated to the plane wave in the forward direction for
longer crystals at optimal focusing.  \refeqnn{Eqn:twophotonAmplitude}
shows that the intensity will be $\propto w_{0\tp}^2$, because, at
optimal focusing we have $z_R = L/\xi_p$, where $z_R$ is given by
\refeqn{Eqn:Rayleigh}, and thus $w_{0\tp}^2 \propto L$. The total flux
is now $\propto L\sqrt{L}$.

As a last step we include filtering. In the previous steps we looked
at a single frequency of the emission, which means that the bandwidth
was narrow enough for the emission to be a single-mode (at optimal
focusing). For narrow enough bandwidths we therefore get a flux
\begin{align}
  P \propto L\sqrt{L} \Delta\lambda_\text{narrow},
\end{align}
which is valid both in free-space and in fiber. As an effect of the
phase-matching conditions there will be a tight connection between the
spectral and spatial modes, as we described in Section
\ref{Sec:singleCoupling} for frequency filtering. In terms of
fiber-coupling this means that when the fiber spatially filters the
emission it will also effectively do frequency filtering. The
bandwidth of the signal emission (810 nm) coupled into single-mode
fibers (using no separate frequency filter) is given by
\begin{align}
  \Delta\lambda_\text{SM} = B/L,
  \label{Eqn:SMbandwith}
\end{align}
where the value $B = 1.23 \times 10^{-11}$ $[\text{m}^2]$ is found for
\mbox{PPKTP} when both the pump and fiber are focused optimally, see
\reffig{Fig:bandwidth_vs_crystal_length}.  We will refer to this
bandwidth as the single-mode bandwidth. It will also determine how
narrow the bandwidth of a filter ($\Delta\lambda_\text{narrow} <
\Delta\lambda_\text{SM}$) need to be for any given length of the
crystal to be considered narrow. The photon flux in the fiber will be
\begin{align}
  P \propto L\sqrt{L} \Delta\lambda_\text{SM} = \sqrt{L},
\end{align}
for any filter ${\Delta\lambda > \Delta\lambda_\text{SM}}$. In
\reffig{Fig:flux_vs_crystal_length} we have plotted the flux for
different filters, $\Delta\lambda_\text{narrow} <
\Delta\lambda_\text{SM} < \Delta\lambda_\text{wide}$.  For filter
bandwidths that are ``wide enough,'' $\Delta\lambda_\text{wide}$, the
free-space emission will be multimode even at optimal pump focusing,
and the free-space photon flux becomes
\begin{align}
  P \propto \sqrt{L}\ g(\Delta\lambda_\text{wide}),
\end{align}
where $g$ is some unknown and non-trivial function determined by the
properties of the crystal material via the Sellmeier equations.

These results clearly show that it is advantageous to have long
crystals as the photon-rate will always monotonically increase even
when coupling the emission into single-mode fibers. As an effect, we
can keep the pump power low, promoting the use of a compact and cheap
laser. This requires that we change the focusing of both the pump
$\xi_\tp$ and the fibers $\xi_{\ts,\ti}$ to the optimal for some
length $L$. Additionally, longer crystals give narrower bandwidth,
which is very advantageous in many applications of entangled photons.
For example, in time-multiplexed schemes it is crucial that the photon
packets keep their widths in the fibers and do not broaden due to
chromatic dispersion, and the broadening can be limited by having a
narrow bandwidth. Another way of reducing the effect of broadening is
by introducing negative dispersion using an appropriatly designed
fiber Bragg grating.  In general these have to be custom manufactured
for broad bandwidths, but for telecom bandwidths, 30-80 GHz, (in the
C-band, between 1525-1562 nm) these are standard off-the-shelf items,
and corresponds to wavelength bandwidths of about 0.25-0.65 nm at 1550
nm. We can see from \refeqn{Eqn:SMbandwith} that ${ 70-180 \text{\ 
    mm}}$ long crystals are needed, taking into account the conversion
factor between signal and idler bandwidths [$\Delta\lambda_\ti =
(\lambda_{0\ti}/\lambda_{0\tp}-1)^2 \Delta\lambda_\ts \approx 3.66
\times \Delta\lambda_\ts$].  Narrow bandwidth can of course be
obtained by the use of spectral filters, however, our results show
that it is better in terms of photon-rates to use long crystals to
achieve small bandwidths rather than to strongly filter the emission
of a short crystal. (This is in contrast to what is claimed by Lee
\textit{et al.} in Ref.  \cite{LEW04}, for birefringent phase-matching
and intersecting cones.)  Furthermore, with narrow bandwidth follows
also long coherence length of the photons which is highly desirable
when working with interferometry as is commonly done when using
time-multiplexing analyzers to code and decode qubits.

\subsection{\label{Sec:M2Coupling}$M^2$ and coupling}
In this subsection we will present the numerical predictions of the
emission mode in terms of the beam quality factor $M^2$ for different
focusing conditions. We will also elaborate on the connection between
the beam quality factor and the coupling efficiency.

\begin{figure}[tb!]
  \begin{center}
    \includegraphics[scale = 0.4]{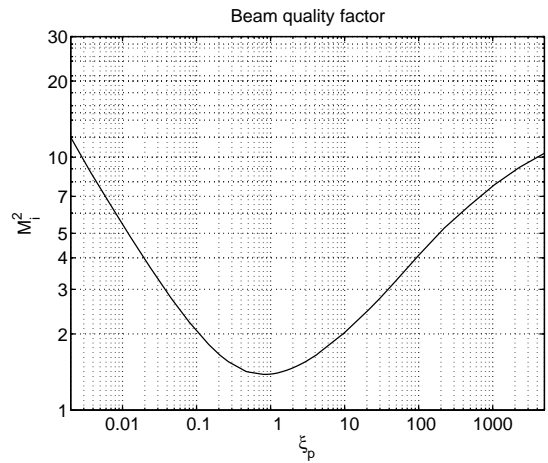}
    \caption{The beam quality factor $M^2$ of the idler plotted
      against the pump beam focusing $\xi_\tp$. The smallest value,
      $M^2 = 1.4$, is found for $\xi_\tp = 0.9$}
    \label{Fig:M2_vs_focusing}
  \end{center}
\end{figure}

\reffigg{Fig:M2_vs_focusing} shows the beam quality factor $M_\ti^2$
plotted against the focusing of the pump for a narrow enough frequency
bandwidth of the idler emission ($\Delta\lambda_\text{narrow} \ll
\Delta\lambda_\text{SM}$). There is a clear optimal focusing, where
the emission reaches close to single-mode, $M_\ti^2 = 1.4$, at a
focusing of $\xi_\tp = 0.9$.  These results are valid for any length
of the crystal, compare to \reffig{Fig:coupling_vs_crystal_length}. A
low value of $M^2$ means that the light is close to a single-mode, and
thus possible to couple well into a single-mode fiber. For bandwidths
larger than the single-mode bandwidth $\Delta\lambda_\text{wide} \gg
\Delta\lambda_\text{SM}$, the light will become spatially multimode
and the coupling efficiency will decrease accordingly.

\begin{figure}[tb!]
  \begin{center}
    \includegraphics[scale = 0.4]{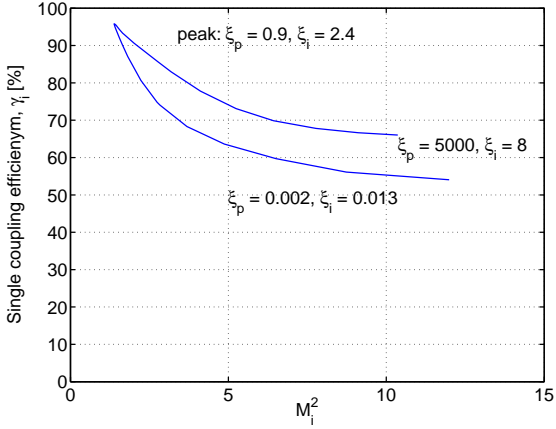}
    \caption{The single coupling $\gamma_\ti$ versus the $M^2$ of the
      idler, using the same data as in \reffig{Fig:M2_vs_focusing} and
      \reffig{Fig:coupling_vs_signalidler}. The graph is parametrized
      by the pump beam focusing and illustrate how a low $M^2$ is
      connected with a large $\gamma_\ti$.}
    \label{Fig:coupling_vs_M2}
    \end{center}
\end{figure}

\reffigg{Fig:coupling_vs_M2} shows the relation between the coupling
efficiency $\gamma_i$ and the $M_\ti^2$, as the focusing $\xi_\tp$ of
the pump is varied. The correspondence is clear, and we can see that
different $M^2$ values can provide the same coupling efficiency. This
is so because the coupling efficiency is only determined by how much
of the emission is in the fundamental mode. What determines the $M^2$
is the distribution of the light between the higher order modes, and
this can differ from one case to another, even with the same amount
contributing to the fundamental mode. In general, as we have said, too
weak focusing will provide spatial transverse multimode, and too
strong focusing will provide spatial longitudinal multimode. It can be
deduced from \reffig{Fig:coupling_vs_M2} that longitudinal multimode,
originating from too strong focusing, creates emission with relatively
higher contribution to the fundamental mode for the same $M^2$ value.

\section{Experimental results}
\label{Sec:experimental}
\begin{figure}[tb!]
  \begin{center}
    \includegraphics[scale = 0.8]{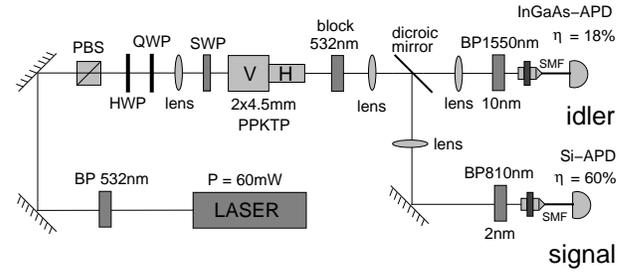}
    \caption{The experimental setup used to create polarization
      entangled photon pairs, and to verify numerical results. PBS:
      polarizing beam splitter; HWP: half wave plate; QWP: quarter
      wave plate; SWP: short-pass filter; BP: band-pass filter; SMF:
      single-mode fiber; $\eta$: detection efficiency.}
    \label{Fig:experimentalsetup}
  \end{center}
\end{figure}

To verify some of the numerical results we compared with experiments.
We have measured the beam quality factor, the bandwidth in the fiber,
and the coupling efficiencies for different focusing conditions of the
pump. The experimental setup is shown in
\reffig{Fig:experimentalsetup}. As a pump we use a frequency doubled
YAG laser emitting approximately 60 mW in the TEM$_{00}$ mode at 532
nm.  Its $M_\tp^2$--value was measured to 1.06.  After a band-pass
filter (BP532), which removes any remaining infrared light, we "clean
up" the polarization using a polarizing beam splitter (PBS).  The
polarization is controlled by a half wave plate (HWP) and a quarter
wave plate (QWP) in front of the crystal. The pump-beam is focused
onto the crystal using a achromatic doublet lens ($f_\tp = 50\ 
\text{mm}$) which introduces a minimal amount of aberrations not to
destroy the low $M^2$ value. The QWP is set to undo any polarization
elliptisation effects caused by the lens, and fluorescence caused by
the same lens is removed by a Schott filter (KG5).

The next component is the crystal. This is a periodically poled, bulk
4.5 mm long KTP crystal, with a poling period of $\Lambda = 9.6\ 
\mu\text{m}$, which will colinearly create a signal at 810 nm and an
idler at 1550 nm when heated in an oven to a temperature $T \approx
100^\circ$.  When the setup is used to create polarization
entanglement, two crystals are present, one oriented for V and one for
H, and the polarization of the pump is set to $45^\circ$. By coupling
the emission from both crystals into single-mode fibers we cannot even
in principle determine which crystal the photons came from, except by
their polarization degree of freedom, and therefore the signal and
idler will interfere in the diagonal basis and get entangled in
polarization.  This principle was first demonstrated by Kwiat
\textit{et al.} in Ref.  \cite{KWWAE99}. Our first results was
presented in Ref.  \cite{PMLTKFCL04}, and the latest results,
overcoming some problems of crystal dispersion and using optimal
focusing, will be found in Ref. \cite{LT05b}.

After the crystal, we block the pump light by a 532 nm band-stop
filter, and the signal and idler emission is focused by achromatic
doublet lenses. The rather small $F$-number ($F = f/D$, where $f$ is
the focal length and $D$ is the beam diameter) of the emitted light
($F < 40$ for $f_\tp = 50 \text{\ mm}$ and $F < 9$ for $f_\tp = 12
\text{\ mm}$) requires good quality lenses not to increase the
$M^2$ factor. The lenses we use are all aberration free down to $F
\approx 6\!-\!11$, and are also quite insensitive to an offset in the
alignment of the optical axis.

To determine the coupling efficiencies and bandwidths, the complete
setup of \reffig{Fig:experimentalsetup} was used. To separate the 810
nm and 1550 nm emission we used a dichroic mirror made for a
$45^\circ$ angle of incidence. The first lens ($f_{\ts\ti} = 30\ 
\text{mm}$) is common to both signal and idler and its task is to
refocus the beams somewhere near the dichroic mirror. The next two
lenses ($f_\ts = 60\ \text{mm}$ and $f_\ti = 40\ \text{mm}$) collimate
each beam, and they are focused into the fiber-tips (with the mode
field diameters being $\text{MFD}_{810} = 5.5\ \mu\text{m}$ and
$\text{MFD}_{1550} = 10.4\ \mu\text{m}$) using aspherical lenses with
$f = 11 \text{\ mm}$.  In front of the fiber couplers we have first
Schott filters (RG715) to block any remaining pump light, and then
interference filters of 2 nm and 10 nm at the 810 nm and 1550 nm side
respectively (BP). The detectors used were a Si-based APD (PerkinElmer
SPCM-AQR-14) for 810 nm and a homemade InGaAs-APD (Epitaxx) module for
1550 nm.

When determining the beam quality factor, $M^2$, we used only a single
crystal oriented to create vertical (V) polarized light, and the
complete setup of \reffig{Fig:experimentalsetup} was also not used.
Instead, we focused the idler emission directly using a lens of focal
length $f_\ti = 75\ \text{mm}$ placed at a distance of 75 mm from the
V-crystal to collimate the beam. At the additional distance of 470 mm
we placed another lens with focal length $f_\ti = 150\ \text{mm}$ that
refocused the beam again, so that we could take measurements of the
beam profile around its waist.

\subsection{\label{Sec:expM2Results}$M^2$ measurements, results}
\begin{figure}[tb!]
    \begin{center}
      \includegraphics[scale = 0.4]{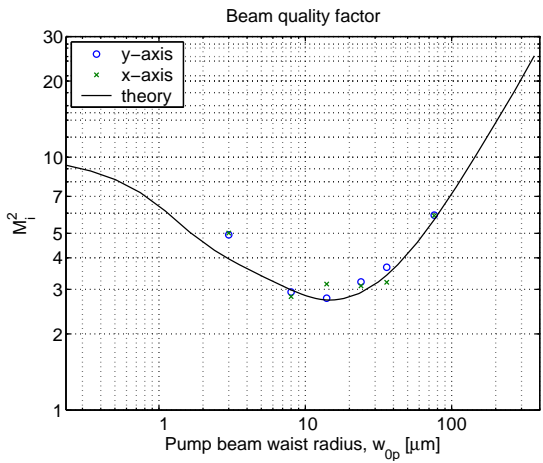}
    \caption{The experimentally observed beam quality factor,
      $M_\ti^2$, for the idler beam at different sizes of the pump
      beam waist radius $w_{0\tp}$. The lowest value of the $M_\ti^2$
      is 2.8 at a $14\ \mu\text{m}$ pump waist.}
    \label{Fig:M2_focus}
    \end{center}
\end{figure}

To obtain the results of \reffig{Fig:M2_focus} we first took images of
the refocused idler beam in the $x$-$y$ plane using an InGaAs-detector
camera from Indigo Systems, model Alpha NIR.  
Several images were acquired for different positions along the
$z$-axis around the waist, and we then integrated the resulting
2-dimensional surface over one axis to create an intensity profile for
the remaining axis.  Because of the detector noise we could not use
the standard deviation method to find the beam radius, defined by the
$1/e^2$ level. Instead we matched a Gaussian shaped function to the
intensity profile to find the width.  This is accurate enough for
mode-shapes that are close to Gaussian, which is the case for low
$M^2$ values.  To limit the impact of the noise we applied a function
that assigned greater weight to the center-values of the intensity
profile. The widths of the beam for each $z$-axis position were then
set together to find the beam profile of the emission, and its $M^2$
factor was determined by fitting to the standard Gaussian-beam
function, \refeqn{Eqn:Gaussian_beam}. We now repeated the procedure
for different focal lengths, $f_\tp$, of the pump lens: 12 mm, 30 mm,
50 mm, 75 mm, 100 mm, and 150 mm, each being placed at a distance that
set the focus in the center of the crystal.  The result, which is
shown in \reffig{Fig:M2_focus}, agrees fairly well with the numerical
predictions.  The shortest focal length lens, 12 mm, gave a somewhat
higher $M^2$, which can be explained by the fact that this was the
only singlet lens used, probably adding some aberrations, while the
others where achromatic doublets.  The lowest value, $M_\ti^2 = 2.8$,
was found with the 50 mm lens giving a $14\ \mu\text{m}$ pump waist
radius $w_{0\tp}$ inside the crystal, corresponding to $\xi_\tp = 2$
for the 4.5 mm long V-crystal (for later reference we observe that
$\xi_\tp = 1.3$ for $L = 3\ \text{mm}$ agrees a little bit better with
numerical results). Note that the $M^2$ values are slightly higher
here compared to \reffig{Fig:M2_vs_focusing}. This can be explained by
the non-perfect phase-matching in the experimental case, resulting
from either too low crystal temperature, uncertainty in the true value
of the poling period (possibly deviating somewhat from its
specification), or both.

\subsection{\label{Sec:expCoupling}Coupling efficiencies, results}
The experimental data for the coupling efficiencies were obtained with
the source producing polarization-entanglement using two crystals. For
this reason we expect the values to be a bit lower than predicted as
we needed to focus the fiber-matched modes for both the H and the V
crystal at the same time. We also have this problem with the pump
beam, and we aimed at placing the focus at the intersecting faces of
the two crystals for both the pump and the fiber. As already
mentioned, the temperature of the crystal used in the experiment was
set lower than required for absolute perfect phase-matching at 810 nm
and 1550 nm. This was because we observed higher photon fluxes at this
setting. Contradictory as it may seem, the explanation is that the
peak of the emission spectrum is not symmetrically centered around the
above wavelengths, but rather towards $810-\alpha$ and $1550+\beta$,
including a long tail representing the emission at larger angles. As
our filters are centered for 810 nm and 1550 nm, the peaks of the
emission can be moved to line up with these by changing the
temperature, and thus the phase-matching, which will give somewhat
higher fluxes although the coupling efficiencies will decrease
according to our definitions.  In addition to having a slightly wrong
poling period these effects degrades the efficiencies, which we could
verify numerically and which is supported by comparing
\reffig{Fig:M2_focus} and \reffig{Fig:M2_vs_focusing}.  The obtained
results for the single coupling efficiencies were $\gamma_s = 32\%$
and $\gamma_i = 79\%$, for the conditional coincidence $\mu_{\ti |
  \ts} = 34\%$, and for the pair coupling $\gamma_c = 11\%$, when
focusing according to $\xi_\tp = 2.1$, $\xi_\ts = 3.2$, and $\xi_\ti =
2.5$ (as decided by available lenses, and assuming $L = 4.5$ mm). For
these numbers we have compensated for the $35\%$ transmission of the
1550 nm filter, and the $85\%$ transmission of the 810 nm filter. The
singles photon rate in the signal fiber was 2.3 Mcps ($10^6$
counts/sec) and in the idler fiber 2.4 Mcps. The total generated rate
of photons before fiber coupling was estimated at 8.6 Mcps and the
coincidence rate in the fibers was 274 kcps, (see Ref. \cite{LT05b}).

\subsection{\label{Sec:expBandwidth}Bandwidth, results}
We have used a spectrograph (SpectraPro 500i, ARC) to measure the
bandwidth of the signal emission using the single-mode fiber without a
filter. The bandwidth was 4 nm for the V-crystal and 6 nm for the
H-crystal.  \reffigg{Fig:bandwidth_vs_crystal_length} suggests that
the effective length of the crystal being poled must be 3 mm and 2 mm
respectively.  Also, from \reffig{Fig:flux_vs_crystal_length}, for the
2 nm filter, we can deduce that the 2 mm crystal should give roughly
$55\%$ of the photon rate of that of the 3 mm one.  Experimental
agreement is good, as we saw the H-crystal giving half the rate of the
V-crystal (with no compensation done by balancing the fiber coupling
or rotating the pump polarization). Referring again to
\reffig{Fig:M2_focus} using the effective crystal length, the best
pump beam focusing parameter is modified to ${\xi_\tp = 1.3}$ for ${L
  = 3\ \text{mm}}$ (V-crystal) which agrees roughly with the value of
optimal focusing, $\xi_\tp = 0.9$.

\section{Concluding discussion}
\label{Sec:conclusions}
In summary, precise focusing of the pump-beam and the fiber-matched
modes can significantly increase the coupling and coincidence
efficiencies of quasi-phase matched SPDC-sources, which is important
for applications needing highly correlated pairs of single photons to
propagate in fibers. We have shown how the beam quality factor of the
emission changes with the focusing of the pump.  At optimal focusing
the emission is mostly created in a spatial single-mode, which couples
well into single-mode fibers, and by maintaining a fixed geometry of
the beam profile in relation to different lengths of the crystal this
stays true for all lengths. We have also shown how the photon flux
depends on the crystal length for different frequency filters, the
conclusion being that longer crystals produce more photons per unit
time at a smaller bandwidth.

In all of the calculations we have assumed a monochromatic (CW) pump
laser. Looking for a possible extension to pulsed operation we observe
that the interaction time, $T$, in \refeqn{Eqn:stateSphere} for a CW
laser is set by the coherence time of the pump alone, and as $T$ is
infinite it transforms into a delta-function of frequency in
\refeqn{Eqn:stateIntegrals}. Using pulsed light, the integral
$\int_{0}^{T} \exp(-i\Delta\omega t)$ should be replaced by
$\int_{-\infty}^{\infty} h(t) \exp(-i\Delta\omega t)$, where $h(t)$ is
the convolution, $h(t) = h_C(t)*h_L(t)$, between the form of the
temporal wave-packet of the pump, $h_C(t)$, and the form of the
crystal along the $z$-axis, $h_L(t)$. We observe that when $h_C(t)$ is
narrow, like for pulsed operation, the transform of $h(t)$ will
instead become a sinc-function, specifying an inexact energy-matching
condition.  Preliminary numerical calculations then show increased
$M^2$ values and decreased coupling efficiencies. However, due to the
characteristics of the convolution, it seems we can retain the good
results of CW even for pulsed operation by using very long crystals,
as this will bring back the delta-function at the limit of infinitely
long crystals. For this discussion we have not yet worried about any
dispersion effects that might come with long crystals and short pump
pulses.

\begin{acknowledgments}
  We would like to thank G. Björk and A. Karlsson for their valuable
  comments and suggestions throughout the work, M.  Pelton and P.
  Marsden for their initial work on the source, A.  Fragemann, C.
  Canalias, and F.  Laurell for providing us with crystals, and J.
  Waldebäck for his skills with electronics.  This work was supported
  by the Swedish Foundation for Strategic Research (\mbox{SSF}) and by
  the European Commission through the integrated project \mbox{SECOQC}
  (Contract No. IST-2003-506813).
\end{acknowledgments}

\appendix
\section{The two-photon frequency and angular spectral amplitude}
\label{App:two-photonAmplitude}
The evolution of the number state vector is given by
\begin{align} 
  \ket{\psi} &= \exp\left[-i\frac{1}{\hbar} \! \int \limits_{t_0}^{t_0
      + T}\!\!  \ud t\ 
    \hat H(t)\right]\ket{\psi_{00}}\nonumber\\*
  & \approx \left(\openone + \frac{1}{i\hbar} \! \int
    \limits_{t_0}^{t_0 + T}\!\! \ud t\ \hat
    H(t)\right)\ket{\psi_{00}},
  \label{Eqn:evolution}
\end{align}
where $\ket{\psi_{00}}$ is the state at time $t_0$, $T$ is the time of
interaction, and $\hat H(t)$ is the Hamiltonian
\begin{align}
  \hat H(t) = \int \limits_{V}\!  \chi^{(2)} \hat E_\tp^{(+)}\hat
  E_\ts^{(-)}\hat E_\ti^{(-)} \ud^3r\ + \text{H.c.}
  \label{Eqn:hamiltonian}
\end{align}
There are three interacting fields in the crystal's volume $V$
ignoring all higher-order terms ($n \geq 3$) of the nonlinearity
$\chi^{(n)}$. All three fields have the same polarization ($ZZZ$):

\begin{subequations}
  \label{Eqn:fields}
    \begin{align}
      & E_\tp^{(+)} \!= \sum \limits_{\boldsymbol{s}_\tp}
      A_\tp(\boldsymbol{s}_\tp) e^{i(k_\tp \boldsymbol{s}_\tp \cdot
        \boldsymbol{r} -
        \omega_\tp t + \phi_\tp)}, \\*
      & \hat E_\ts^{(-)} \!=\! \int\!\!\ud \phi_\ts \!\! \int\!\!\ud
      \omega_\ts A(\omega_\ts) \sum \limits_{\boldsymbol{s}_\ts}
      e^{-i(k_\ts \boldsymbol{s}_\ts \cdot \boldsymbol{r} - \omega_\ts
        t + \phi_\ts)}\hat{a}^\dagger_\ts(\omega_\ts,
      \boldsymbol{s}_\ts), \\*
      & \hat E_\ti^{(-)} \!=\! \int\!\!\ud \phi_\ti \!\! \int\!\!\ud
      \omega_\ti A(\omega_\ti) \sum \limits_{\boldsymbol{s}_\ti}
      e^{-i(k_\ti \boldsymbol{s}_\ti \cdot \boldsymbol{r} - \omega_\ti
        t + \phi_\ti)}\hat{a}^\dagger_\ti(\omega_\ti,
      \boldsymbol{s}_\ti).
    \end{align}
\end{subequations}
The field of the pump is classical and monochromatic so that we can
replace $\hat E_\tp^{(+)} $ by $ E_\tp^{(+)}$. The plus-sign denotes
conjugation, i.e. annihilation (+) or creation (-) of the state. In all
the calculations we use the notation $\boldsymbol{k} = k
\boldsymbol{s}$, where $\boldsymbol{s}$ is the unit length vector of
$\boldsymbol{k}$.  The angular amplitude spectrum
$A_\tp(\boldsymbol{s}_\tp)$ takes into account the focusing of the
pump.  For signal and idler, we sum over both frequency and angular
modes, where $\hat{a}(\omega, \boldsymbol{s})$ is the field operator,
and $A(\omega)$ is the frequency amplitude of a Gaussian shaped
detector filter having the bandwidth $\Delta\lambda$ ({\small FWHM})
and center wavelength $\lambda_c^\ph$ (all wavelengths in vacuum). Via
the relation $\omega = 2\pi c n_{\!  \scriptscriptstyle \lambda} /
\lambda$ its form is given by
\begin{align} 
  A(\omega; \lambda) = e^{-2\log(2)(\lambda - \lambda_c^\ph)^2 /
    \Delta\lambda^2 }.
\end{align}
Each signal and idler photon is created with a random phase,
$\phi_\ts$ and $\phi_\ti$ respectively, which we also need to sum
over. The only nonzero solution is completely correlated phases as
will be shown later. The phase of the pump $\phi_\tp$ is constant but
arbitrary.

For periodically poled materials, the nonlinearity $\chi^{(2)}$ has
sharp boundaries, and later on in the calculations it will facilitate
to make an expansion of $\chi^{(2)}$ into its Fourier-series
components
\begin{align} 
  \chi^{(2)} = \chi_2^\ph\ f(\boldsymbol{r}) = \chi_2^\ph \sum_{m =
    0}^\infty f_m e^{-i m \boldsymbol{K} \cdot \boldsymbol{r}},
\end{align}
and then do a sinusoidal approximation using the first term,
\begin{align}
  \chi^{(2)} = \chi_2^\ph\ f_1 e^{-i \boldsymbol{K}\cdot
    \boldsymbol{r}},
  \label{Eqn:chi2}
\end{align}
where $\boldsymbol{K} = {2 \pi / \Lambda}\ \boldsymbol{e}_z$, and $
\Lambda$ is the grating period. Appendix \ref{App:expansion} treats
the case of a $M+1$ term series expansion.

From \refeqn{Eqn:evolution} the number state becomes
\begin{align} 
  \ket{\psi} &= \ket{\psi_{00}} + \int\!\!\!\int\!\!\ud \omega_\ts \ud
  \omega_\ti \sum\limits_{\boldsymbol{s}_\ts}\sum
  \limits_{\boldsymbol{s}_\ti} S(\omega_\ts, \omega_\ti,
  \boldsymbol{s}_\ts, \boldsymbol{s}_\ti)
  \hat{a}^\dagger_\ts \hat{a}^\dagger_\ti \ket{\psi_{00}} \nonumber\\*
  & = \ket{\psi_{00}} + G_2 \ket{\psi_{11}},
\end{align}
where $G_2$ is the unnormalized amplitude for the two-photon
number state,
\begin{align} 
  G_2 = \langle{\psi_{11}}\ket{\psi} = \int\!\!\!\int\!\!\ud
  \omega_\ts \ud \omega_\ti \sum\limits_{\boldsymbol{s}_\ts}\sum
  \limits_{\boldsymbol{s}_\ti} S(\omega_\ts, \omega_\ti,
  \boldsymbol{s}_\ts, \boldsymbol{s}_\ti),
  \label{Eqn:G2Short}
\end{align}
such that for $t_0 = 0$,
\begin{align} 
  \frac{1}{i\hbar} \int \limits_0^T\! \ud t\ \hat H(t) &= G_2
  \ \hat{a}^\dagger_\ts \hat{a}^\dagger_\ti - \text{H.c.}.
  \label{Eqn:intHamiltonian}
\end{align}

Our goal now is to arrive at an expression for the amplitude $S$ which
will also enter in the state of frequency and angular spectrum of the
form
\begin{align} 
  \ket{\psi_{\omega, \boldsymbol{s}}} = \int\!\!\!\int\!\!\ud
  \omega_\ts \ud \omega_\ti \sum\limits_{\boldsymbol{s}_\ts}\sum
  \limits_{\boldsymbol{s}_\ti} S(\omega_\ts, \omega_\ti,
  \boldsymbol{s}_\ts, \boldsymbol{s}_\ti) \ket{\omega_\ts}
  \ket{\omega_\ti} \ket{\boldsymbol{s}_\ts} \ket{\boldsymbol{s}_\ti}.
  \label{Eqn:stateForm}
\end{align}
We start by inserting \refeqn{Eqn:chi2} into \refeqn{Eqn:hamiltonian}
and then \refeqn{Eqn:hamiltonian} into \refeqn{Eqn:intHamiltonian}
which gives
\begin{align}
  G_2 &= \frac{1}{i\hbar} \int \limits_0^T\! \ud t \int
  \limits_{V}\!\! \ud^3r\ \chi_2^\ph\ f_1 e^{-i \boldsymbol{K}\cdot
    \boldsymbol{r}} E_\tp^{(+)} E_\ts^{(-)} E_\ti^{(-)}.
\label{Eqn:G2Expanded}
\end{align}
By making a substitution of the fields in \refeqn{Eqn:fields} into
\refeqn{Eqn:G2Expanded}, and via identification using
\refeqn{Eqn:G2Short} we find that\\

\noindent $S(\omega_\ts, \omega_\ti, \boldsymbol{s}_\ts,
\boldsymbol{s}_\ti) = \nonumber$
\begin{align}
  & \quad\ \chi_2^\ph\ f_1 A(\omega_\ts)A(\omega_\ti) \sum
  \limits_{\boldsymbol{s}_\tp} A_\tp(\boldsymbol{s}_\tp)
  \nonumber\\*
  & \times\ \int \limits_{-L/2}^{L/2}\!\!\!\! \ud z\!\! \int
  \limits_{-\infty}^{\infty}\!\!\! \ud y\!\! \int
  \limits_{-\infty}^{\infty}\!\!\! \ud x\ e^{-i \Delta
    \boldsymbol{k}\cdot (x \boldsymbol{e}_x + y
    \boldsymbol{e}_y + z \boldsymbol{e}_z)} \nonumber\\*
  & \times\ \frac{1}{i\hbar} \iint \limits_0^{\quad 2\pi}\! \ud
  \phi_\ts \ud \phi_\ti \int \limits_0^T\! \ud t\ e^{-i[(\omega_\ts +
    \omega_\ti - \omega_\tp)t\ +\ \phi_\ts + \phi_\ti - \phi_\tp]},
\end{align}
where the volume integral has been expressed in a Cartesian coordinate
system ($\boldsymbol{r} = x\boldsymbol{e}_x + y\boldsymbol{e}_y +
z\boldsymbol{e}_z$, see \reffig{Fig:coordinate}),
\begin{align}
  \int \limits_{V}\! \ud^3r\ = \int \limits_{-L/2}^{L/2}\!\!\!\! \ud
  z\!\! \int \limits_{-\infty}^{\infty}\!\!\! \ud y\!\! \int
  \limits_{-\infty}^{\infty}\!\!\! \ud x.
\end{align}
We have also introduced the phase mismatching vector
\begin{subequations}
  \label{Eqn:mismatchVector}
  \begin{align}
    \Delta \boldsymbol{k} & = k_\ts \boldsymbol{s}_\ts + k_\ti
    \boldsymbol{s}_\ti - k_\tp
    \boldsymbol{s}_\tp + \boldsymbol{K} \\
    & = \Delta k_x \boldsymbol{e}_x + \Delta k_y \boldsymbol{e}_y +
    \Delta k_z \boldsymbol{e}_z.
  \end{align}
\end{subequations}
In a Cartesian coordinate system the normalized vectors
$\boldsymbol{s}$ are represented by
\begin{align}
  \boldsymbol{s}_\ts &= p_\ts\ \boldsymbol{e}_x + q_\ts\ 
  \boldsymbol{e}_y +
  m_\ts\ \boldsymbol{e}_z, \nonumber\\
  \boldsymbol{s}_\ti &= p_\ti\ \boldsymbol{e}_x + q_\ti\ 
  \boldsymbol{e}_y +
  m_\ti\ \boldsymbol{e}_z, \nonumber\\
  \boldsymbol{s}_\tp &= p_\tp\ \boldsymbol{e}_x + q_\tp\ 
  \boldsymbol{e}_y +
  m_\tp\ \boldsymbol{e}_z, \nonumber\\
  \boldsymbol{K} &= K\ \boldsymbol{e}_z,
\end{align}
where $p$, $q$, and $m$ are the normalized components of
$\boldsymbol{s}$ in each of the three dimensions \cite{MW95}.

Because of the rotational symmetry of the emitted modes, it is
suitable to use a spherical coordinate system $(\theta, \varphi)$, for
which $p = \sin\theta\cos\varphi,\ q = \sin\theta\sin\varphi,$ and $m
= \cos\theta$. The phase-mismatch vector components then become
\begin{align}
  \Delta k_x &= k_\ts \sin\theta_\ts\cos\varphi_\ts + k_\ti
  \sin\theta_\ti\cos\varphi_\ti - k_\tp \sin\theta_\tp\cos\varphi_\tp,
  \nonumber\\
  \Delta k_y &= k_\ts \sin\theta_\ts\sin\varphi_\ts + k_\ti
  \sin\theta_\ti\sin\varphi_\ti - k_\tp \sin\theta_\tp\sin\varphi_\tp,
  \nonumber\\
  \Delta k_z &= k_\ts \cos\theta_\ts + k_\ti \cos\theta_\ti - k_\tp
  \cos\theta_\tp + K.
  \label{Eqn:mismatchVectorComponents}
\end{align}
Note that the magnitude of the signal and idler \mbox{$k$-vectors}
implicitly depends on the polar angle $\theta$ according to
\begin{subequations}
  \label{Eqn:kvectorAngleDependence}
  \begin{align}
    k_\ts(\theta_\ts) &= 1 /
    \sqrt{\left(\frac{\cos\theta_\ts}{k_\ts^Z}\right)^2 +
      \left(\frac{\sin\theta_\ts}{k_\ts^Y}\right)^2}, \\
    k_\ti(\theta_\ti) &= 1 /
    \sqrt{\left(\frac{\cos\theta_\ti}{k_\ti^Z}\right)^2 +
      \left(\frac{\sin\theta_\ti}{k_\ti^Y}\right)^2},
  \end{align}
\end{subequations}
where $k_\ts^Z$, $k_\ts^Y$, $k_\ti^Z$, and $k_\ti^Y$ are the constant
magnitude of the $k$-vectors along the crystals $Z$ and $Y$ axis,
respectively ($k_\tp$ need to be constant and equal to $k_\tp^Z$ as we
will soon show).  Generally, there is negligible difference in
refractive indices between the crystal's $X$ and $Y$ axes which
cancels the dependence on the azimuthal angle $\varphi$ in the
equations above. We therefore use the $Y$ axis as the major axis being
orthogonal to $Z$.

Using spherical coordinates exclusively leads to\\

\noindent $S(\omega_\ts, \omega_\ti, \theta_\ts, \theta_\ti, \varphi_\ts,
\varphi_\ti) = \nonumber$
\begin{align}
  & \quad\ \chi_2^\ph\ f_1 A(\omega_\ts) A(\omega_\ti) \int
  \limits_{0}^{\pi/2}\!\! \sin\theta_\tp\ \ud \theta_\tp\!\! \int
  \limits_{0}^{2\pi}\!\!
  \ud \varphi_\tp\ A_\tp(\theta_\tp, \varphi_\tp) \nonumber\\*
  & \times\ \int \limits_{-L/2}^{L/2}\!\!\!\! \ud z\!\! \int
  \limits_{-\infty}^{\infty}\!\!\! \ud y\!\! \int
  \limits_{-\infty}^{\infty}\!\!\! \ud x\ 
  e^{-i [\Delta k_x x + \Delta k_y y + \Delta k_z z]} \nonumber\\*
  & \times\ \frac{1}{i\hbar} \iint \limits_0^{\quad 2\pi}\! \ud
  \phi_\ts \ud \phi_\ti \int \limits_0^T\! \ud t\ e^{-i[(\omega_\ts +
    \omega_\ti - \omega_\tp)t\ +\ \phi_\ts + \phi_\ti - \phi_\tp]}.
  \label{Eqn:stateSphere}
\end{align}
The angular spectral amplitude $A_\tp$ of the pump beam in
\refeqn{Eqn:stateSphere} is Gaussian shaped for a laser emitting in a
TEM$_{00}$ single-mode, and in spherical coordinates it becomes
\cite{MW95}
\begin{align}
  A_\tp(\theta_\tp, \varphi_\tp) &= \frac{k_\tp
    w_{0\tp}^\ph}{\sqrt{2\pi}} e^{-(k_\tp w_{0\tp}^\ph)^2
    \sin^2\theta_\tp/4},
  \label{Eqn:angularSpectrum}
\end{align}
where the beam waist radius $w_{0\tp}^\ph$ of the focused pump-beam
has entered the calculations. The function is normalized to represent
the same constant power available in the beam at different focusing
conditions.

Now we will solve the integrals over space, time, and phase in
\refeqn{Eqn:stateSphere}. In doing so we note that there are three
spatial integrals of which two are the Fourier transforms of unity
($\ud x$ and $\ud y$) and one is the transform of a box-function ($\ud
z$).  The transforms turn into two $\delta$-functions and a
sinc-function respectively. The time-integral also turns into a
$\delta$-function of the three frequencies $\omega_\ts$, $\omega_\ti$,
and $\omega_\tp$. This is because we have a monochromatic pump-beam
with infinite coherence length, which effectively leads to an infinite
interaction-time, ${T \rightarrow \infty}$, even for short crystals.
The two integrals over the random phases $\phi_\ts$ and $\phi_\ti$
will make the amplitude $S$ vanish completely if the phases are not
fully correlated with each other. Therefore, the only nonzero
solution is when the two phases add up to a constant.  $S$ can be
complex-valued, thus yielding the relation ${\phi_\ts + \phi_\ti =
  \phi_\tp + C}$. If we
let $C = 0$ for simplicity, we are led to\\

\noindent $S(\omega_\ts, \omega_\ti, \theta_\ts, \theta_\ti,
\varphi_\ts, \varphi_\ti) = \nonumber$
\begin{align}
  & \quad\ \chi_2^\ph\ f_1 A(\omega_\ts) A(\omega_\ti) \int
  \limits_{0}^{\pi/2} \!\! \sin\theta_\tp\ \ud \theta_\tp\!\! \int
  \limits_{0}^{2\pi}\!\!
  \ud \varphi_\tp\ A_\tp(\theta_\tp, \varphi_\tp) \nonumber\\*
  & \times\ \delta(\Delta k_x)\ \delta(\Delta k_y)\ 
  L\ \text{sinc}\left[\frac{L}{2}\Delta k_z\right] \nonumber\\*
  & \times\ \frac{4\pi^2}{i\hbar} \delta(\omega_\ts + \omega_\ti -
  \omega_\tp).
  \label{Eqn:stateIntegrals}
\end{align}

We now have two integrals over $\theta_\tp$ and $\varphi_\tp$ with
$\delta$-functions over $\Delta k_x$ and $\Delta k_y$ which in turn
depends on $\theta_\tp$ and $\varphi_\tp$ according to
\refeqn{Eqn:mismatchVectorComponents}. The integrals can be canceled
in a few steps by setting the equalities ${\Delta k_x = 0}$ and
${\Delta k_y = 0}$, and to that end we need to assume that $k_\tp$ is
constant for small angles $\theta_\tp$, i.e. ${k_\tp = k_\tp^Z}$ which
we believe is a fair approximation for pump-light that is not
extremely focused. By extreme we mean beyond the validity of the
paraxial approximation. The latter equality applied to
\refeqn{Eqn:mismatchVectorComponents} gives
\begin{align}
  \varphi_\tp^\prime = \arcsin\left(\frac{k_\ts
      \sin\theta_\ts\sin\varphi_\ts + k_\ti
      \sin\theta_\ti\sin\varphi_\ti}{k_\tp^Z\sin\theta_\tp^\prime}\right).
  \label{Eqn:phiExplicit}
\end{align}
\refeqnn{Eqn:phiExplicit} together with the relation $\arcsin(x) =
\arccos(\sqrt{1-x^2})$ now gives the following expression for ${\Delta
  k_x = 0}$ of \refeqn{Eqn:mismatchVectorComponents} (with
$\varphi_\tp$ primed),
\begin{align}
  & k_\ts \sin\theta_\ts\cos\varphi_\ts + k_\ti
  \sin\theta_\ti\cos\varphi_\ti
  + \nonumber \\
  & - k_\tp^Z \sin\theta_\tp^\prime\sqrt{1-\left(\frac{k_\ts
        \sin\theta_\ts\sin\varphi_\ts + k_\ti
        \sin\theta_\ti\sin\varphi_\ti}{k_\tp^Z\sin\theta_\tp^\prime}\right)^2}
  = 0.
  \label{Eqn:thetaImplicit}
\end{align}
If we now take the square of \refeqn{Eqn:thetaImplicit} and solve for
$\theta_\tp^\prime$ we get
\begin{align}
  \theta_\tp^\prime = \arcsin\sqrt{P^2 + Q^2} = \arccos\sqrt{1- (P^2 +
    Q^2)},
  \label{Eqn:thetaExplicit}
\end{align}
where
\begin{subequations}
  \label{Eqn:PandQ}
  \begin{align}
    P &= \frac{k_\ts \sin\theta_\ts\sin\varphi_\ts + k_\ti
      \sin\theta_\ti\sin\varphi_\ti
     }{k_\tp^Z}, \\
    Q &= \frac{k_\ts \sin\theta_\ts\cos\varphi_\ts + k_\ti
      \sin\theta_\ti\cos\varphi_\ti}{k_\tp^Z}.
  \end{align}
\end{subequations}
Furthermore,
\begin{align}
  P^2 + Q^2 &= \nonumber \\
  & \frac{k_\ts^2 \sin^2\theta_\ts + k_\ti^2 \sin^2\theta_\ti + 2
    k_\ts k_\ti \sin\theta_\ts \sin\theta_\ti
    \cos(\delphi)}{(k_\tp^Z)^2},
  \label{Eqn:PandQSquared}
\end{align}
where we are allowed to introduce ${\delphi = \varphi_\ts -
  \varphi_\ti}$.  This is a result of the assumption of rotational
symmetry and will lead to the final state being invariant to a common
variation in the azimuthal angles for signal, $\varphi_\ts$, and
idler, $\varphi_\ti$.  As shown here, only the angle-difference is of
importance.  Using \refeqn{Eqn:thetaExplicit} in the expression for
$\Delta k_z$ of \refeqn{Eqn:mismatchVectorComponents} we have
\begin{align}
  \Delta k_z^\prime &= k_\ts \cos\theta_\ts + k_\ti \cos\theta_\ti -
  k_\tp^Z \sqrt{1 - (P^2 + Q^2)} + K.
  \label{Eqn:deltakPrime}
\end{align}

At this stage the two integrals in \refeqn{Eqn:stateIntegrals} have
been canceled and the amplitude can be simplified as\\

\noindent $S(\omega_\ts, \omega_\ti, \theta_\ts, \theta_\ti, \delphi) =
\nonumber$
\begin{align}
  & \quad\ \chi_2^\ph\ f_1 A(\omega_\ts) A(\omega_\ti)
  A_\tp(\theta_\tp^\prime,
  \varphi_\tp^\prime) \nonumber\\*
  & \times\ 
  L\ \text{sinc}\left[\frac{L}{2}\Delta k_z^\prime\right] \nonumber\\*
  & \times\ \frac{4\pi^2}{i\hbar} \delta(\omega_\ts + \omega_\ti -
  \omega_\tp).
\end{align}

One further simplification includes the observation that the frequency
$\delta$-function can be reduced to unity by introducing a common
frequency $\epsilon$ instead of $\omega_\ts$ and $\omega_\ti$ as
defined by ${\omega_\ts = \omega_{0\ts} + \epsilon},\ {\omega_\ti =
  \omega_{0\ti} - \epsilon}$, so that for two matched filters the form
of the filter amplitude becomes squared.  Using also
\refeqn{Eqn:thetaExplicit} together with \refeqn{Eqn:angularSpectrum}
the expression for the amplitude of the state of frequency and angular
spectrum finally becomes\\

\noindent $S(\epsilon, \theta_\ts, \theta_\ti, \delphi) = \nonumber$
\begin{align}
  & \quad\ \frac{4\pi^2 \chi_2^\ph f_1 L}{i\hbar} A^2(\epsilon)
  \frac{k_\tp^Z w_{0\tp}^\ph}{\sqrt{2\pi}} e^{-(k_\tp^Z
    w_{0\tp}^\ph)^2 [P^2 + Q^2]/4} \nonumber\\*
  & \times \text{sinc}\left[\frac{L}{2}\left(k_\ts\cos\theta_\ts +
      k_\ti\cos\theta_\ti - k_\tp^Z\sqrt{1 - (P^2 + Q^2)} + K
    \right)\right],
  \label{Eqn:stateFinal}
\end{align}
where $P^2 + Q^2$ is defined by \refeqn{Eqn:PandQSquared} and
the $k_\ts$'s and $k_\ti$'s by \refeqn{Eqn:kvectorAngleDependence}.

We now have a final expression for the two-photon amplitude
\begin{align} 
  G_2 = \int\!\! \ud \epsilon \int\!\!\! \int\!\! \sin\theta_\ts \ud
  \theta_\ts \sin\theta_\ti \ud \theta_\ti\! \int\!\! \ud \delphi\ 
  S(\epsilon, \theta_\ts, \theta_\ti, \delphi),
\end{align}
which gives the two-photon state-vector in terms of frequency and
angular spectrum in the form of \refeqn{Eqn:stateForm}
\begin{align} 
  \ket{\psi_{\epsilon, \theta, \delphi}} = G_2 \ket{\epsilon}
  \ket{\theta_\ts} \ket{\theta_\ti} \ket{\delphi}.
\end{align}

\section{Series expansion of \large{$\chi^{(2)}$}}
\label{App:expansion}
The poling structure of periodically poled crystal has the approximate
form of a square-function along the $z$-axis. In such a case, the
$M+1$ term series expansion of $\chi^{(2)}$ become
\begin{align} 
  \chi^{(2)} = \chi_2^\ph\ f(\boldsymbol{r}) = \frac{4
    \chi_2^\ph}{\pi} \sum_{m = 0}^M \frac{(-1)^m}{2 m + 1} e^{-i (2m
    +1) \boldsymbol{K} \cdot \boldsymbol{r}},
  \label{Eqn:chiExpanded}
\end{align}
where $\boldsymbol{K} = {2 \pi / \Lambda}\ \boldsymbol{e}_z$, and $
\Lambda$ is the grating period.  In the following expression we have
isolated the $z$-dependent part of \refeqn{Eqn:stateSphere}:
\begin{align} 
  {\chi_2^\ph f_1}\!\! \int \limits_{-L/2}^{L/2}\!\! \ud z\ e^{-i \Delta
    k_z z}.
\end{align}
Now, putting the series expansion of $\chi^{(2)}$ into the
calculations of Appendix \ref{App:two-photonAmplitude}, the former
expression should be replaced by
\begin{align} 
  \frac{4 \chi_2^\ph}{\pi}\!\! \int \limits_{-L/2}^{L/2}\!\! \ud z
  \sum_{m = 0}^M \frac{(-1)^m}{2 m + 1} e^{-i \Delta k_z^{(m)} z},
  \label{Eqn:stateExpansion}
\end{align}
where
\begin{align} 
 \Delta k_z^{(m)} &= \Delta k_z^{\prime} + 2 m K.
\end{align}
By reversing the order of the sum and the integral in
\refeqn{Eqn:stateExpansion} we can identify a Fourier transform of
box-function with an extra phase.  The result of the transform is a
sinc, providing thus
\begin{align} 
  \frac{4 \chi_2^\ph}{\pi} \sum_{m = 0}^M \frac{(-1)^m}{2 m + 1}
  \text{sinc}\left[\frac{L}{2} (\Delta k_z^{\prime} + 2mK)\right], 
\end{align}
which is the final expression to replace the sinc-function in the
state amplitude, \refeqn{Eqn:stateFinal}, having now $M+1$ terms to
approximate the square-shaped poling structure.  For $M = 0$ the
expression reduces to the sinusoidal approximation with $f_1 = 4/\pi$.

\bibliography{ljunggren}

\end{document}